\documentclass[a4paper]{article}

\usepackage{amssymb, amsmath, amsthm}
\usepackage{commath}
\usepackage[squaren]{SIunits}
\usepackage[lofdepth,lotdepth,caption=false]{subfig}

\usepackage{fp}
\usepackage{pgfplots}
\usepackage{tikz}
\usetikzlibrary{fixedpointarithmetic}
\usetikzlibrary{external}
\usepackage[numbers]{natbib}
\usepackage{accsupp} 

\usepackage[margin=2.0cm]{geometry}

\usetikzlibrary{arrows}
\usetikzlibrary{calc, decorations.pathmorphing}
\usepgfplotslibrary{groupplots}
\pgfplotsset{compat=1.5}
\pgfplotsset{tick scale binop=\times}

\renewcommand{\vec}{\boldsymbol}

\newlength\figureheight
\newlength\figurewidth
\definecolor{color1}{rgb}{0.75,0,0.75}
\definecolor{color0}{rgb}{0,0.75,0.75}
\definecolor{color2}{rgb}{0.75,0.75,0}

\tikzset{myblue/.style={blue, solid, thick}}
\tikzset{mydblue/.style={blue, dotted, thick}}
\tikzset{myred/.style={red, densely dotted, thick}}
\tikzset{myblack/.style={black, dashed, thick}}
\tikzset{mygreen/.style={green!50!black, densely dashed, thick}}
\tikzset{mycyan/.style={color0, dashdotted, thick}}
\tikzset{mypurple/.style={color1, loosely dashdotted, thick}}
\tikzset{myyellow/.style={color2, dashdotdotted, thick}}
\tikzset{myorange/.style={orange, loosely dashed, thick}}

\usepackage{titlesec}
\titleformat*{\section}{\bf\large}
\titleformat*{\subsection}{\bf\normalsize}
\titleformat*{\subsubsection}{\itshape}
\titleformat*{\paragraph}{\itshape}

\title{\bf Automatic calibration of damping layers in finite element time
  domain simulations}

\author{Steven Vandekerckhove\thanks{Wave Propagation and Signal Processing
    Research Group, KU Leuven - Kulak, Etienne Sabbelaan 53, 8500 Kortrijk,
    Belgium}
    \and Garth N. Wells\thanks{Department of Engineering, University of
      Cambridge, Trumpington Street, Cambridge CB2 1PZ, United Kingdom}
    \and Herbert De Gersem\thanks{Institut f\"ur Elektromagnetischer Felder,
      Technische Universit\"at Darmstadt, Schlossgartenstrasse 8,
      64289 Darmstadt. Germany}
    \and Koen Van Den Abeele\thanks{Wave Propagation and Signal Processing
      Research Group, KU Leuven - Kulak, Etienne Sabbelaan 53, 8500 Kortrijk,
      Belgium}}

\date{}
\begin{document}

\maketitle

\begin{abstract}
  \noindent Matched layers are commonly used in numerical simulations
  of wave propagation to model (semi-)infinite domains.  Attenuation
  functions describe the damping in layers, and provide a matching of
  the wave impedance at the interface between the domain of interest
  and the absorbing region. Selecting parameters in the attenuation
  functions is non-trivial. In this work, an optimisation procedure
  for automatically calibrating matched layers is presented. The
  procedure is based on solving optimisation problems constrained by
  partial differential equations with polynomial and
  piecewise-constant attenuation functions. We show experimentally
  that, for finite element time domain simulations, piecewise-constant
  attenuation function are at least as efficient as quadratic
  attenuation functions. This observation leads us to introduce
  consecutive matched layers as an alternative to perfectly matched
  layers, which can easily be employed for problems with arbitrary
  geometries. Moreover, the use of consecutive matched layers leads to
  a reduction in computational cost compared to perfectly matched
  layers.  Examples are presented for acoustic, elastodynamic and
  electromagnetic problems.  Numerical simulations are performed with
  the libraries FEniCS/DOLFIN and dolfin-adjoint, and the computer
  code to reproduce all numerical examples is made freely available.
\end{abstract}

\maketitle
\section{Introduction}

Three types of boundary conditions are frequently used in numerical
wave propagation problems: reflecting boundaries, modelled by
homogeneous Dirichlet and Neumann conditions; ports through which
energy enters or leaves the system, modelled by non-homogeneous
Dirichlet or Neumann boundary conditions; and boundary conditions that
mimic open space when truncating an infinite domain. A number of
strategies for truncating infinite domains have been developed,
including absorbing boundary conditions~\citep{hagstrom99,
  peterson88}, absorbing layers~\citep{holland83williams, katz76etall}
and one-way approximations~\citep{engquist77majda, mur81}.  An
absorbing layer introduces damping and is realised by extending the
computational domain beyond the domain of interest, and it is
desirable to keep the size of the absorbing domain as small as
possible to limit the additional computational work.  However, none of
the early damping layer techniques proved to be flawless.

In 1994, \citet{berenger94} introduced an absorbing domain called
Perfectly Matched Layers (PMLs).  In a PML, waves are damped at a
certain rate, described by an attenuation function (AF). It is
desirable to use an `optimal' AF in order to limit the size of the
PML.  Unfortunately, there is no universal recipe available to
determine the best AF for specific problems. For particular cases,
optimal PMLs can be found through mathematical analysis. For example,
\citet{chew96jin} proved that for finite difference time domain
methods, second-order polynomial AFs are optimal and suggested that
these results should also be expected for finite element time domain
methods. A generalisation of the analysis to more complicated cases
(unstructured meshes, more general geometries and loads) is not
straightforward, may be suboptimal or may even fail.

In this work we present an automatic calibration procedure for PMLs
through optimisation of the PML parameters for a given problem. The
functional we attempt to minimise is the energy left in the domain
after an input signal should have left the domain of interest. The
problem is constrained by the considered differential equation that
describes the wave propagation of interest.  We use gradient-based
optimisation procedures to determine the parameters, with the adjoint
of the forward problem used to compute the derivative of the target
functional with respect to the PML parameters.  We consider polynomial
and piecewise-constant AFs, with the latter case motivating the
introduction of what we will call `Consecutive Matched Layers'
(CMLs). An advantage of CMLs is that they are easily added to problems
with arbitrary geometries, as we will show through numerical examples.

Numerical examples of the proposed procedure are presented for
acoustic, elastodynamic and electromagnetic problems. The examples use
the FEniCS/DOLFIN~\citep{logg10wells, logg12etall, alnaes:2015} and
dolfin-adjoint~\citep{farrell13etall, funke13farrell} libraries. The
complete source code to produce the presented examples is freely
available and provided as supporting material~\citep{pmlcode}.

The remainder of the paper is organised as follows. The considered
wave propagation problems are described in Section~\ref{sec:problem},
followed by the introductions of PMLs in Section~\ref{sec:pml}. In
Section~\ref{sec:cml}, the formulation of Consecutive Matched Layers
(CMLs) is presented, which is followed by the proposed procedure for
automatic calibration of PMLs and CMLs in Section~\ref{sec:setup}. We
present and discuss test cases and results in
Section~\ref{sec:results}. Conclusions are drawn in
Section~\ref{sec:conclusions}.

\section{Wave propagation problems}
\label{sec:problem}

We will consider acoustic, elastodynamic and electromagnetic wave
propagation problems. Each of these problems is defined in this
section, but we first present a generic formulation in which these
problems can framed.

On a spatial domain $\Omega \subset \mathbb{R}^{d}$, where $1 \le d
\le 3$, we consider linear wave propagation problems in the generic
form
\begin{equation}
  \label{eq:genericwave}
  \Dot{\vec{q}} + \sum_{i = 1}^{d} \vec{F}_{i, i}
    = \vec{f} \qquad \text{on} \ \Omega \times[0,\, T],
\end{equation}
where $\vec{q}$ is a vector of length $n$ containing the $n$ unknown
fields, $\vec{F}_{i} = \vec{A}_i \vec{q}$ is a flux vector of
length~$n$, $\vec{f}$~is a source function of length $n$ and $T$ is
the final time. The matrices $\vec{A}_i$ contain material parameters
and will be defined for each specific problem we consider. The
notation $\vec{F}_{i,i} = \partial \vec{F}_{i}/ \partial x_{i}$ (no
summation) implies component-wise partial differentiation of
$\boldsymbol{F}_{i}$ with respect to~$x_{i}$.  Boundary conditions
will be presented later for each specific problem.

The first considered model is acoustic wave propagation, described by
the system
\begin{equation}
  \begin{aligned}
    \frac{1}{K}\Dot{p} &=  -\nabla \cdot \vec{v},
    \\
    \rho \Dot{\vec{v}} &=  -\nabla p + \vec{f},
  \end{aligned}
  \label{eq:asyst}
\end{equation}
where $K > 0$ is the bulk modulus, $p$ is the pressure, $\vec{v}$ is
the velocity, $\rho > 0$ is the mass density and $\vec{f}$ is an
applied body force. This problem is transformed into the generic
form~\eqref{eq:genericwave}, in three-dimensions, with $\vec{q} =
(v_1, v_2, v_3, p)^T$ and the matrices $\vec{A}_i$
in~\eqref{eq:Aacoustic}.

The second model concerns electromagnetic wave propagation, described
by the system
\begin{equation}
  \begin{aligned}
    \mu \Dot{\vec{H}}  &=  -\nabla \times \vec{E},
    \\
    \varepsilon \Dot{\vec{E}}  &=  \nabla \times \vec{H} - \vec{J},
  \end{aligned}
  \label{eq:emsyst}
\end{equation}
where $\mu > 0$ is the permeability, $\vec{H}$ is the magnetic field
strength, $\vec{E}$ is the electric field strength, $\varepsilon > 0$
is the permittivity and $\vec{J}$ is a current density. This problem
is transformed into the generic form~\eqref{eq:genericwave}, in three
dimensions, with $\vec{q} = (H_1, H_2, H_3, E_1, E_2, E_3)^T$ and the
matrices $\vec{A}_i$ in~\eqref{eq:Aelectro}.

Finally, linearised elastic wave propagation will be considered, and
is described by the system
\begin{equation}
  \begin{aligned}
    \vec{\mathcal{C}}^{-1} : \Dot{\vec{T}}
    &= \frac{1}{2}\del{\nabla\vec{v} + \del{\nabla\vec{v}}^T},
    \\
    \rho \Dot{\vec{v}} &= \nabla \cdot \vec{T}+ \vec{f},
  \end{aligned}
  \label{eq:esyst}
\end{equation}
where $\vec{\mathcal{C}}$ is the fourth-order, isotropic elastic
stiffness tensor, $\vec{T}$ is the stress tensor, $\vec{v}$ is the
particle velocity, $\rho > 0$ the mass density and $\vec{f}$ is an
applied body force. This problem is transformed into the generic
form~\eqref{eq:genericwave}, in three dimensions, with $\vec{q} =
\del{v_1, v_2, v_3, T_{11}, T_{22}, T_{33}, 2T_{23}, 2T_{13},
  2T_{12}}^T$ and the matrices $\vec{A}_i$ in~\eqref{eq:Aelasto}.

\section{Perfectly matched layers}
\label{sec:pml}

We denote the domain of physical interest by $\Omega_i$, which is
extended with an absorbing domain~$\Omega_a$ ($\Omega_i \cap \Omega_a
= \emptyset$), leading to the computational domain~$\Omega =
\Omega_i \cup \Omega_a$.  To obtain a formulation for wave propagation
problems with PMLs, we apply the technique of complex coordinate
stretching \citep{chew94weedon, teixeira00chew} to the generic wave
equation in~\eqref{eq:genericwave}.

Solutions to wave equations are of the form
\begin{equation}
  \vec{q}(\vec{x},t) = \Bar{\vec{q}}\del{\vec{x}}e^{-\jmath \omega t},
  \\
  \label{eq:fd}
\end{equation}
where $\Bar{\vec{q}}(\vec{x})$ is the spatial solution and $\omega$ is
the frequency. A frequency domain formulation can be used by noting
that~$\Dot{\vec{q}} = -\jmath \omega \vec{q}$. PMLs in all directions
are applied by introducing the coordinate transformations
\begin{equation}
  \pd{}{x_i}
  \rightarrow
  \del{\frac{1}{1+\jmath\frac{\sigma_i(x_i)}{\omega}}}
  \pd{}{x_i},
  \label{eq:pml3d}
\end{equation}
where $\sigma_i(x_i)$ are attenuation functions (AFs), and which are
non-zero only in the absorbing region~$\Omega_a$. The AFs will be
defined at the end of the section.

We will denote combinations of different AFs in the index,
e.g.~$\sigma_{ij + k} = \sigma_i \sigma_j + \sigma_k$.
Using~\eqref{eq:fd} and applying the coordinate transformations
in~\eqref{eq:pml3d} to the wave equation~\eqref{eq:genericwave} leads
to
\begin{equation}
  -\jmath\omega \vec{q}
  + \sum_{i = 1}^{d} \frac{\vec{F}_{i, i}}{1 -
  \frac{\sigma_{i}}{\jmath\omega}}
  = \vec{f}.
  \label{eq:addpml}
\end{equation}
Multiplying \eqref{eq:addpml} by all denominators appearing in it
leads to
\begin{equation}
  -\jmath \omega \del{\prod_{i = 1}^{d}\del{1 -
  \frac{\sigma_{i}}{\jmath\omega}}} \vec{q}
  + \sum_{i=1}^{d}\del{\prod_{\substack{j = 1
        \\ j \neq i}}^{d}
  \del{1 - \frac{\sigma_{j}}{\jmath\omega}}}\vec{F}_{i, i}
  = \del{\prod_{i=1}^{d}\del{1 - \frac{\sigma_{i}}{\jmath\omega}}} \vec{f}.
  \label{eq:addpml2}
\end{equation}
With no source term inside the absorbing region $\Omega_a$, we have
$\sigma_{i}\vec{f} = \vec{0}$ and the right-hand side
of~\eqref{eq:addpml2} simplifies to~$\vec{f}$.  Expanding the
remaining products leads to
\begin{multline}
  -\jmath\omega \vec{q}
  + \sum_{i = 0}^{3} \vec{F}_{i, i}
  + \sigma_{1 + 2 + 3}\vec{q}
  - \frac{1}{\jmath\omega}\sigma_{12+13+23}\vec{q}
  + \frac{1}{\del{\jmath\omega}^2}\sigma_{123}\vec{q}
  - \frac{1}{\jmath\omega} \sum_{i = 0}^{3} \del{\sum_{\substack{j = 1 \\ j
  \neq i}}^d\sigma_{j}}\vec{F}_{i,i}
\\
  + \frac{1}{\del{\jmath\omega}^2}\sum_{i = 0}^{3} \del{\prod_{\substack{j =
  1 \\ j \neq i}}^d\sigma_{j}}\vec{F}_{i,i}
  = \vec{f}.
  \label{eq:pmlfreq}
\end{multline}
To obtain a system of first-order equations from~\eqref{eq:pmlfreq},
for $d = 3 $ two auxiliary fields, $\Dot{\vec{r}} = \vec{q}$ and
$\Dot{\vec{s}} = \vec{r}$ are introduced, resulting in two auxiliary
differential equations (ADEs) in addition to the wave equation:
\begin{equation}
  \begin{aligned}
    \Dot{\vec{q}} + \vec{F}_{1, 1} + \vec{F}_{2, 2} + \vec{F}_{3, 3}
    + \sigma_{1+2+3}\vec{q} +\vec{r}
    &= \vec{f},
    \\
    \Dot{\vec{r}} - \sigma_{12+13+23}\vec{q} - \sigma_{2+3}\vec{F}_{1,1}
    - \sigma_{1+3}\vec{F}_{2,2} - \sigma_{1+2}\vec{F}_{3,3} - \vec{s}
    &= \vec{0},
    \\
    \Dot{\vec{s}} + \sigma_{123}\vec{q} + \sigma_{23}\vec{F}_{1,1}
    + \sigma_{13}\vec{F}_{2,2} + \sigma_{12}\vec{F}_{3,3}
    &= \vec{0}.
  \end{aligned}
  \label{eq:fullpml}
\end{equation}
In two spatial dimensions ($d = 2$) , we have the simplified system:
\begin{equation}
  \begin{aligned}
    \Dot{\vec{q}} + \vec{F}_{1, 1} + \vec{F}_{2, 2} +
    \sigma_{1+2}\vec{q} +\vec{r}
    &= \vec{f},
    \\
    \Dot{\vec{r}} - \sigma_2\vec{F}_{1,1} +
    \sigma_1\vec{F}_{2, 2} - \sigma_{12}\vec{q}
    &= \vec{0}.
  \end{aligned}
  \label{eq:2dfullpml}
\end{equation}
In one spatial dimension ($d = 1$), there are no ADEs needed to
describe the PML:
\begin{equation}
  \Dot{\vec{q}} + \vec{F}_{1, 1} + \sigma_1\vec{q} = \vec{f}.
  \label{eq:1dfullpml}
\end{equation}

A specific PML is defined by the AFs $\sigma_i$. The literature,
e.g.~\citep{chew96jin}, generally suggests polynomial AFs. For
axis-aligned rectangular (cuboid) domains, polynomial AFs can be
expressed as
\begin{equation}
  \sigma_{i}(x_{i})
  =
  \begin{cases}
    \displaystyle\sum_{j = 0}^{n}c_{ij} \bar{x}_{i}^{j}
    & \text{if} \ x_{i} \in [a_{(i)0}, a_{(i)0}+w_i]
    \\
    0 &\text{otherwise},
  \end{cases}
  \label{eq:polyaf}
\end{equation}
where $n$ is the order of the polynomial, $c_{ij}$ are the
coefficients of the polynomial, $\Bar{x}_{i} = g(x_{i})$ is an affine
transformation of $x_i$ such that $g(x_{i}) = 0$ on the boundary
between the domain of interest and the absorbing region, and $g(x_{i})
= 1$ on the exterior boundary of the absorbing region, $x_i =
a_{(i)0}$ is the interface between $\Omega_i$ and $\Omega_a$ and $w_i$
is the total width of the PML in the $i$th direction.

We also introduce a description of an AF with $N$ piecewise-constant
AFs of the form
\begin{equation}
  \sigma_i(x_i)
  =
  \begin{cases}
    c_{ij} & \text{if} \ x_{i} \in [a_{(i)j}, a_{(i)j+1}] \
    \forall j = 0 \ldots N - 1,
    \\
    0     & \text{otherwise},
  \end{cases}
  \label{eq:pwcaf}
\end{equation}
where $c_{ij} \ge 0$ are scalar values and $a_{(i)j} = a_{i} +
j(w_{i}/N)$.

\section{Consecutive matched layers}
\label{sec:cml}

\begin{figure}
  \centering
  \subfloat[Adding perfectly matched layers in multiple directions to
    a geometry using complex coordinate stretching leads to
    overlapping regions.
    \label{fig:complexpml_overlap}]{%
    \includegraphics[width=0.45\textwidth]{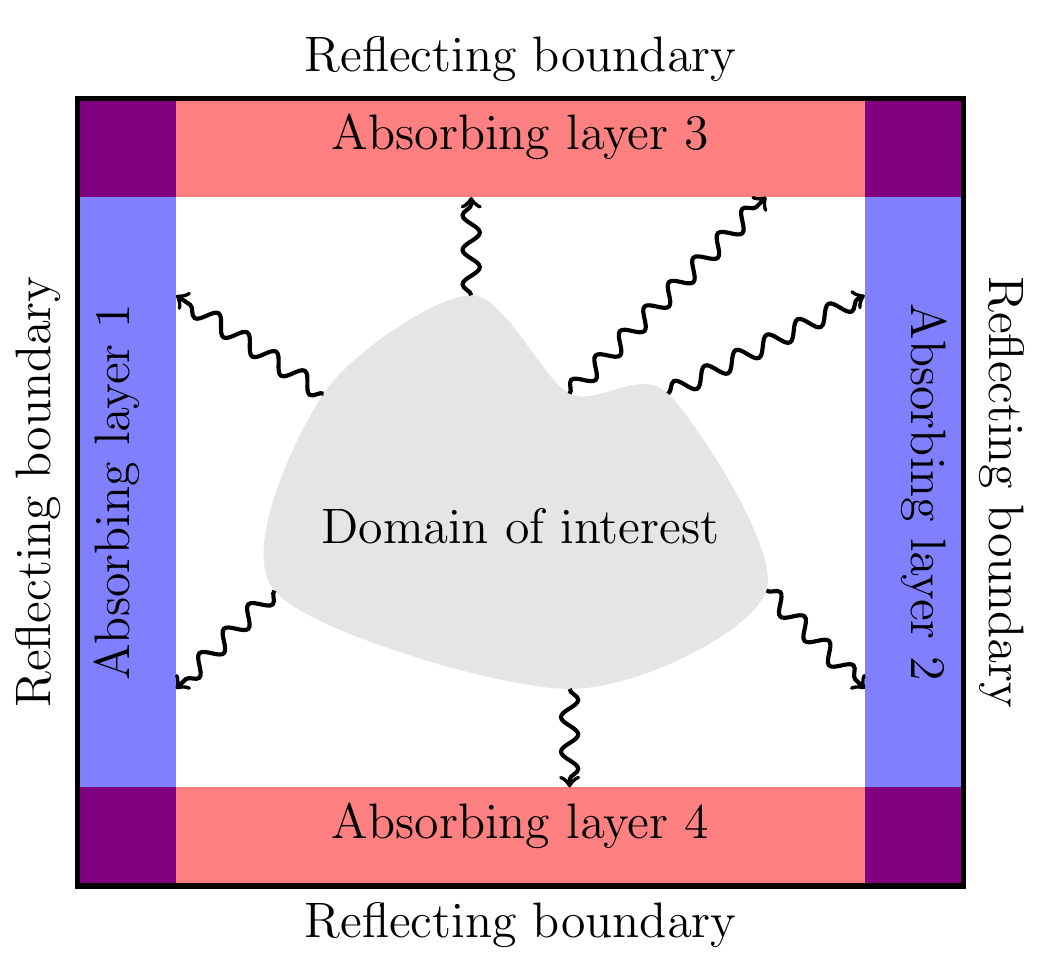}}
  \hfill
  \subfloat[Multiple perfectly matched layers that have been added to
    a geometry without overlap.\label{fig:complexpml_noeverlap}]{%
    \includegraphics[width=0.45\textwidth]{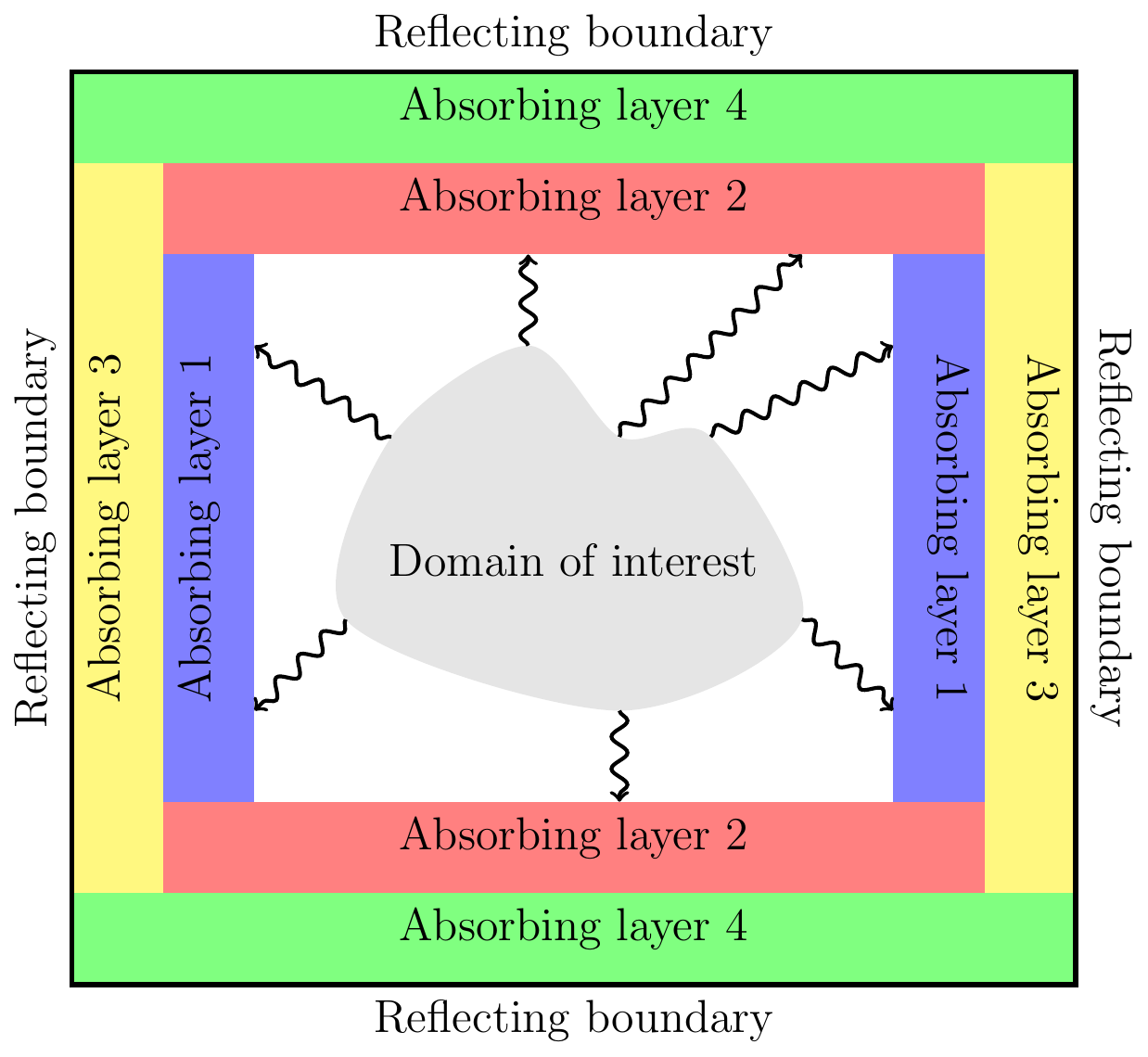}}
  \\
  \subfloat[When two consecutive layers as shown in
    Figure~\ref{fig:complexpml_noeverlap} are defined by the same
    constant value, they can be considered as one merged
    layer. Fragmentation of a matched layer in this manner can be used
    to add tightly wrapped absorbing layers to an arbitrary geometry,
    as demonstrated in Figure~\ref{fig:simplepml}.
    \label{fig:simplepml_square}]{%
    \includegraphics[width=0.45\textwidth]{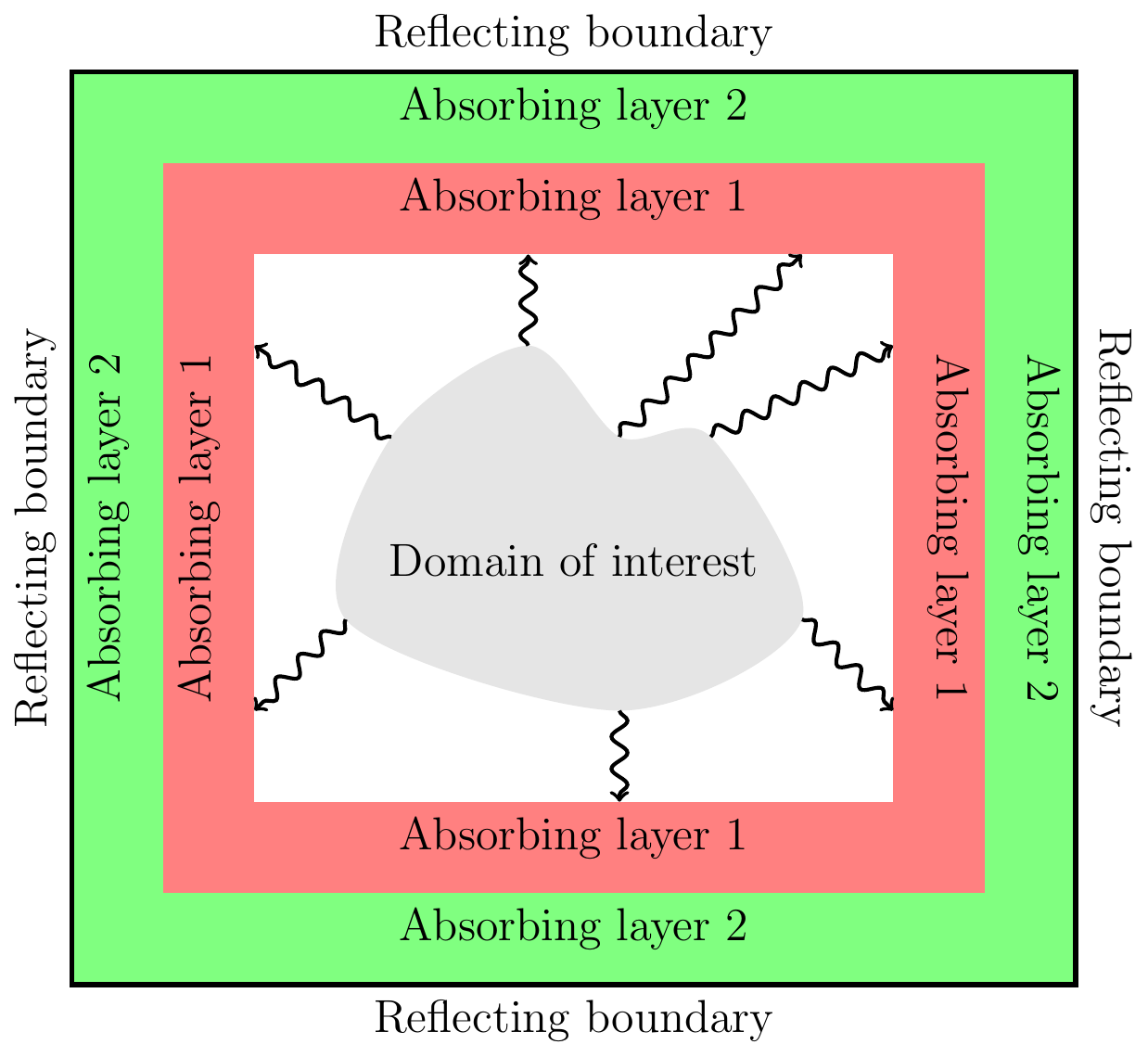}
  }
  \hfill
  \subfloat[Multiple tightly wrapped absorbing layers around an
    arbitrary geometry.\label{fig:simplepml}]{%
    \includegraphics[width=0.45\textwidth]{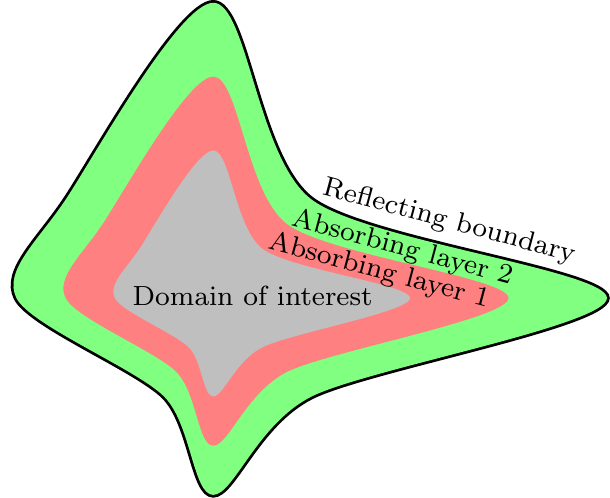}
  }
  \caption{Graphical depiction of perfectly matched layers and
    consecutive matched layers.}
\end{figure}

The complex coordinate stretching procedure used in the previous
section to the PML configuration depicted in
Figure~\ref{fig:complexpml_overlap}. Overlapping PML regions leads to
products of AFs that appear in the ADEs in~\eqref{eq:fullpml}. Solving
for the auxiliary fields adds to the computational cost.  To avoid
this increase in cost, we adopt a simplification to the PML strategy.

When using non-overlapping absorbing domains, as depicted in
Figure~\ref{fig:complexpml_noeverlap}, products of AFs are zero
and~\eqref{eq:fullpml} reduces to
\begin{equation}
  \begin{aligned}
    \Dot{\vec{q}}  + \vec{F}_{1,1} + \vec{F}_{2,2} + \vec{F}_{3,3}
    + \sigma_{1+2+3}\vec{q} +\vec{r} &= \vec{f},
    \\
    \Dot{\vec{r}} - \sigma_{2+3} \vec{F}_{1, 1} - \sigma_{1+3} \vec{F}_{2,2}
    - \sigma_{1+2}\vec{F}_{3, 3} &= \vec{0},
  \end{aligned}
  \label{eq:earlycml}
\end{equation}
which eliminates one ADE compared to~\eqref{eq:fullpml}. If we assume
that $\vec{r} = \vec{0}$, which can be motivated by the fact that
spatial derivatives in the second equation will be relatively small
due to the damping, also the second ADE vanishes, further
reducing~\eqref{eq:fullpml} to
\begin{equation}
  \Dot{\vec{q}} + \vec{F}_{1,1} + \vec{F}_{2,2} + \vec{F}_{3,3}
  + \sigma_{1+2+3}\vec{q} = \vec{f}.
  \label{eq:earlycml2}
\end{equation}
Since we prefer direction-independent AFs, we choose the AF in all
directions to be defined by the same constant. Hence, $\sigma_{1+2+3}$
can be replaced by an AF of the form
\begin{equation}
  \sigma_{i}(\vec{x}),
  =
  \begin{cases}
    c_i & \text{if} \ \vec{x} \in \Omega_{a_i}
    \\
    0     & \text{otherwise},
  \end{cases}
\label{eq:cmlcaf}
\end{equation}
where $c_i \ge 0$ is a constant scalar and $\Omega_{a_i}$ is the $i$th
`layer' of the absorbing domain.  Using the AF in~\eqref{eq:cmlcaf}
for a problem where $\Omega_{a_1} = \Omega_a$ leads to a
simplification of~\eqref{eq:earlycml2}:
\begin{equation}
  \Dot{\vec{q}} + \vec{F}_{1,1} + \vec{F}_{2,2} + \vec{F}_{3,3} +
  \sigma \vec{q} = \vec{f}.
\end{equation}
This formulation closely resembles the original absorbing layer
strategy~\citep{holland83williams}.  We however suggest to consider an
absorbing domain $\Omega_a$ which is divided into $N$ non-overlapping
absorbing layers $\Omega_{a_i}$ such that $\Omega_a = \bigcup_{i=1}^N
\Omega_{a_i}$, which leads to the formulation
\begin{equation}
  \Dot{\vec{q}} + \vec{F}_{1,1} + \vec{F}_{2,2} + \vec{F}_{3,3}
  + \sigma_{i}(\vec{x}) \vec{q}
  = \vec{f},
  \label{eq:cmlaf}
\end{equation}
The resulting configuration is illustrated in
Figure~\ref{fig:simplepml_square}, where the AF is constant on each
colour/layer. Due to the absence of auxiliary fields, the
computational cost is reduced relative to the PML model. A difficulty
is how to choose the terms that define the AFs in the layers. This
issue will be addressed in the following section.

We note that a reasonable domain of interest can be extended with
tightly wrapped layers, as shown in Figure~\ref{fig:simplepml}, and
meshed conformingly. Hence this procedure can be applied to problems
with arbitrary geometries, while avoiding complex mathematical
interventions, e.g., as presented in~\citep{gao08zhang}.

By neglecting the ADE in~\eqref{eq:earlycml} the absorbing domain is
no longer a PML, hence we will refer to the simplified damping
strategy as \emph{consecutive matched layers} (CMLs). This name refers
to the non-reflective free space boundary condition introduced in
\citet{katz76etall}, mentioned as \emph{absorbing layer} in
\citet{holland83williams} and referred to as \emph{matched layer} in
\citet{berenger94}.

\section{Automatic calibration of matched layer problems}
\label{sec:setup}

The matched layer approaches presented in the preceding sections
involved scalar AFs, $\sigma_{i}\del{x_{i}}$, and it is necessary to
define their functional form. For finite difference methods, there are
numerous papers describing how to determine the AFs,
e.g,~\citep{asvadurov03etall, chew96jin,
  collino98monk}. \citet{chew96jin} proved that quadratic polynomials
result in optimal AFs for finite difference methods. Even if we
presume that quadratic polynomial AFs are optimal for finite element
time domain simulations, the question remains what precise form the
quadratic polynomial should take for optimal results. Since users
often want to add damping layers to their models without studying the
truncation strategy in depth, an automatic determination procedure is
appealing. We present a generic recipe for automatic calibration of
the AF coefficients. The presented procedure is based on solving an
optimisation problem.

\subsection{Formulation of the optimisation problem}
\label{sec:opt}

Our abstract optimisation problem is formulated as
\begin{align}
  \min_{\vec{y}, \vec{u}} J(\vec{y}, \vec{u})
  &\qquad \text{(objective functional)},
  \\
  \vec{c}(\vec{y}, \vec{u}) = \vec{0}  &\qquad \text{(constraint)},
\end{align}
where $J$ is a scalar function, $\vec{y}$ is the state vector,
$\vec{u} \in U_\text{ad}$ is the control vector, $U_\text{ad}$ is the
set of the admissible values for the control values and $\vec{c}$
represents a set of constraints. For the considered problems, the
state vector $\vec{y}$ contains $\vec{q}$ in the wave
equation~\eqref{eq:genericwave}, and in the case of PML calibration is
also contains the solutions of the auxiliary fields $\vec{r}$ and
$\vec{s}$ in~\eqref{eq:fullpml}. The control vector $\vec{u}$ contains
the $c_{ij}$ parameters that define the AFs. Satisfaction of the
constraint $\vec{c}(\vec{y}, \vec{u})$ in our case is satisfaction of
the wave equation with matched layers.

The matched layer optimisation problem is in general not convex due to
the non-linear relation between the controls and the states given by
the constraint. This implies the likely existence of multiple local
minima. Consequences of the existence of multiple local minima will be
demonstrated by the numerical examples in
Section~\ref{sec:results}. This prohibits us of considering the
outcome of the automatic calibration procedure as optimal. Based on
experimental results, we will however argue that the outcome is very
likely to have near optimal performance.

\subsection{Measuring the quality of the absorbing region}
\label{sec:quality}

To define an objective functional we need to quantify the quality of a
matched layer problem. Typical quality measures involve the reflection
coefficients, both at the interface between the domain of interest
$\Omega_i$ and the matched layer $\Omega_a$, and within the matched
layer (see~\citep{chew96jin}). This \emph{a priori} quality measure is
difficult to manipulate in combination with finite element
formulations.

We propose quantifying the quality of a matched layer through the
amount of energy in a system at judiciously chosen time for a
judiciously chosen source term.  With reflecting boundaries around the
domain of interest and a vanishing input signal, the total energy in
the system for the considered problems is constant once the input
signal has vanished. If the domain of interest was embedded in an
infinite domain, the total energy in the domain of interest would be
zero at sufficiently large time. When absorbing layers are added to
the domain of interest to mimic an infinite domain, the energy will
reduce over time due to attenuation in the absorbing layers only, but
it is highly unlikely that it will ever be exactly zero. The goal of
the calibration procedure is to choose parameters for the matched
layers such that the energy in the whole computational domain is
minimised at a suitably chosen time, which we will call the
`calibration time',~$T_{c}$.  The reduction in energy in the numerical
simulation at time $T$ due to the absorbing layers is given by
\begin{equation}
  \delta E = -10 \log_{10}
  \del{\frac{E\del{T}}{\bar{E}\del{T}}},
  \label{eq:energyreduction}
\end{equation}
where $\bar{E}$, the energy in the whole computational domain with
zero-valued AFs, is used as a reference value and $E$ is the energy on
the computational domain for the problem with non-zero AFs.

For the problems that we consider, the energy in a system is given by:
\begin{equation}
  E(t) = \frac{1}{2} \left<\vec{Q} \vec{q}(t), \vec{q}(t)
  \right>_\Omega,
  \label{eq:generalenergy}
\end{equation}
where $\left<., . \right>_\Omega$ is the $L_2$ inner product over the
entire computational domain $\Omega$ and $\vec{Q}$ is a matrix
containing the material parameters. Concrete expressions for the
energy and the matrices $\vec{Q}$ for the specific problems in
Section~\ref{sec:problem} are given in~\ref{sec:specifics}.

\subsection{Objective functional}
\label{sec:objective}

The objective functional we use in calibrating matched layer problems is
\begin{equation}
  J(\vec{q}, \vec{u}) = E(T_{c}),
  \label{eq:objective}
\end{equation}
where $E(T_{c})$ is the energy in the system at the calibration time.

Another quantity of interest in designing matched layers is
reflections at the interface between the domain of interest and the
damping region. If the calibration time $T_{c}$ is chosen too large,
then energy can be damped gradually every time a wave encounters the
damping region and is partially reflected by it. In order to include
the effect of these reflections in $J$~\eqref{eq:objective}, the
calibration time should be chosen such that reflections of the input
signal at the material/matched layer interface encounter the damping
region as few times as possible.

A practical concern is that the calibration time should be chosen as
small as possible for computational speed, since a greater calibration
time increases the number of time steps, and hence the cost of the
optimisation process.

\subsection{Computing derivatives of the objective functional}
\label{sec:adjoint}

We will use derivative-based optimisation methods to calibrate the
matched layer parameters. To compute the gradient of the objective
functional $J$ with respect to the control parameters $\vec{u}$, we
use the adjoint approach~\citep{troeltzsch10}. In essence, we find
${\rm d}J / {\rm d} \vec{u}$ from
\begin{equation}
  \od{J}{\vec{u}} = \pd{J}{\vec{u}} - \vec{\lambda}^{T} \od{\vec{c}}{\vec{u}},
\end{equation}
where the adjoint variable $\vec{\lambda}$ is the solution to:
\begin{equation}
  \del{\dpd{\vec{c}}{\vec{q}}}^{T} \vec{\lambda}
    =  \del{\dpd{J}{\vec{q}}}^{T}.
    \label{eq:adjointsystem}
\end{equation}
A detailed derivation for the time discretised problems can be found
in~\ref{sec:gradient}.  Key to the adjoint approach for computing
derivatives of functionals is that that only one system needs to be
solved to compute the gradient, regardless of the number of controls.
Moreover, \eqref{eq:adjointsystem} is similar in structure to the
system that is solved in the forward problem.

For the numerical examples in Section~\ref{sec:results}, in our
implementations we express the forward model in FEniCS
syntax~\citep{alnaes:2014, logg10wells, logg12etall}, from which the
adjoint problem is computed automatically by the library
dolfin-adjoint~\citep{funke13farrell}.

\subsection{Practical procedure}
\label{sec:recipe}

To automatically calibrate a PML or CMLs for a problem of interest we
create a calibration set-up. The procedure is:
\begin{enumerate}
\item Extend the domain of interest with artificial layers $\Omega_a$
  and mesh domain with cell edges conforming to the boundary of
  $\Omega_i$ and~$\Omega_a$.

\item Extend the physical material parameters on the domain of
  interest to the absorbing region.

\item Set the attenuation in the damping region to zero.

\item Select an input signal with local support in time to fit the
  frequency range of the application under consideration.

\item Select a calibration time $T_{c}$, such that the peak of the
  input pulse has travelled at least once through the damping region
  in every direction at the lowest wave speed.

\item Update the AF parameters via a gradient-based optimisation
  process.
\end{enumerate}
When the optimiser has converged, the obtained controls for the
calibration set-up are used in the AF to solve the forward problem of
interest.  Note that the calibration set-up can differ from the
problem of interest, as will be demonstrated for the electromagnetic
example in Section~\ref{sec:results}. In the other example the
geometry, mesh and excitation of the problem of interest and
calibration set-up are kept. The final time of the problem of interest
can differ from the calibration time,~$T_{c}$. We will call the final
time for the problem of interest the `evaluation time',~$T_{e}$.

\section{Numerical examples and discussion}
\label{sec:results}

We present examples using the calibration procedure for finite element
acoustic, elastic and electromagnetic wave propagation problems, and
consider both PMLs and CMLs. We will begin with a one-dimensional
example, before moving on to two- and three-dimensional cases to
examine performance with oblique incidence angles.  We will consider
PMLs for acoustic and elastodynamic examples, and CMLs for
elastodynamic and electromagnetic examples. The computer code for
reproducing all examples is available in the supporting
material~\citep{pmlcode}.

For all examples, we use the L-BFGS-B optimiser from
SciPy~\citep{scipy}. This optimiser is a limited memory BFGS
implementation with bound support~\citep{byrd95etall}. The bound
support is used to prevent the optimiser choosing negative values for
the piecewise-constant AFs. The optimiser stops when the gradient
drops below a chosen threshold~\citep{nocedal00wright}. The threshold
used in the different examples can be found in the supporting
material~\citep{pmlcode}.

To fully define the objective functional in~\eqref{eq:objective}, a
calibration time and input signal have to be chosen. For all examples
we use a Gaussian pulse. We choose the calibration time such that the
peak of the input pulse has time to travel at least once to the
boundary of the computational domain and back to the interface between
the domain of interest and the absorbing domain at the lowest wave
speed. Unless mentioned otherwise, first-order elements are used for
all computations.

\subsection{Perfectly matched layers}
\label{sec:pmlresults}

The examples presented in this section consider polynomial and
piecewise-constant AFs for PMLs, as described in
Section~\ref{sec:pml}.

\subsubsection{Acoustic wave propagation}
\label{sec:acoustic}

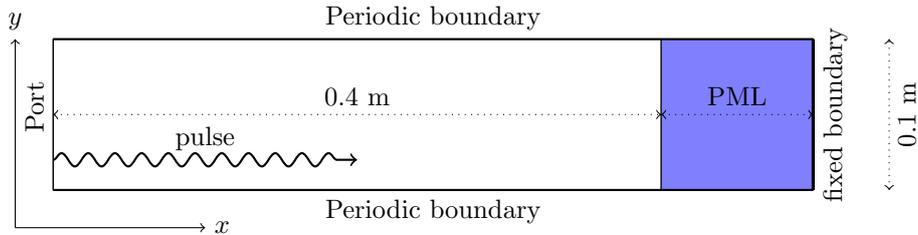
\begin{figure}
  \centering
  \begin{tikzpicture}[scale=20]
    \draw (0, 0) rectangle (0.4, 0.1);
    \draw[fill=blue!50!white] (0.4, 0.1) rectangle (0.5, 0.0);
    \draw (0,0) -- (0, .1)  node
    [above=1mm,midway,rotate=90,anchor=south]
    {Port};
    \draw[thick] (0,0) -- (0.5, 0)  node [below,midway,anchor=north]
    {Periodic
      boundary};
    \draw[thick] (0,.1) -- (0.5, .1)  node [above,midway,anchor=south]
    {Periodic
      boundary};
    \draw[dotted, <->] (0,.05) -- (0.4, .05)  node
    [above,midway,anchor=south]
         {$0.4~\meter$};
         \draw[decorate, decoration={snake, post length=1mm}, ->, thick]
         (0,.02) --
         (0.2, .02)  node [above,midway,anchor=south] {pulse};
         \draw[dotted, <->] (0.4,.05) -- (0.5, .05)  node
         [above,midway,anchor=south]
              {PML};
              \draw[very thick] (0.5,0) -- (0.5, .1)  node
                   [below,midway,rotate=90,anchor=north] {fixed
                   boundary};
                   \draw[dotted, <->] (0.55,0) -- (0.55, .1)  node
                        [below,midway,rotate=90,anchor=north]
                        {$0.1~\meter$};
                        \draw[solid, ->] (-0.025,-0.025) -- (-0.025,
                        0.1)  node
                        [above,anchor=south]
                             {$y$};
                             \draw[solid, ->] (-0.025,-0.025) -- (0.1,
                             -0.025)
                             node [above,anchor=west]
                                  {$x$};
  \end{tikzpicture}
  \caption{Geometry of the two-dimensional acoustic wave example with
    waves propagating in one direction.}
  \label{fig:acoustic}
\end{figure}

We consider a rectangular domain of interest $\Omega_i = [0,
  0.4]~\meter~\times~[0, 0.1]~\meter$, which is extended at the
right-hand boundary with a PML, as depicted in
Figure~\ref{fig:acoustic}. The domain is meshed with crossed-triangle
cells with edge length $0.01~\meter$ in both $x$- and
$y$-directions. Periodic boundary conditions are applied in the
$y$-direction. On the right-hand side of the computational domain, a
reflecting boundary condition with $\vec{v} = \vec{0}$ is applied. An
open boundary on the right-hand side of the domain is modelled by
adding a PML in front of the reflecting boundary.  On the left-hand
boundary, the condition $\vec{v} =
\del{\exp\del{-\del{4(t-t_0)/t_0}^2}, 0}~\meter\per\second$ is
applied, where $t_0$ is the offset for the pulse. Note that for a
large enough time $\vec{v}$ approaches zero and this boundary acts as
a reflecting fixed boundary.
We consider a homogeneous medium with mass density $\rho =
1.269~\kilogram\per\cubic\meter$ and bulk modulus $K =
101000~\pascal$. The time step is 90\% of the CFL condition, $\Delta t
= 0.9 \times 0.01/(v\sqrt{2})~\second$, where $v = \sqrt{K/\rho}$ is
the wave speed for the medium.  The offset of the pulse is chosen to
be $t_0 = 100\Delta t$. The calibration time $T_{c}$ is chosen to be
the time the peak of the pulse needs to travel two and a half times
through the domain of interest, $T_{c} = 2.5(0.4\meter)v + t_{0}$.
This way the peak of the pulse can encounter the PML interface only
once, but there is sufficient time for the pulse to travel
back-and-forth in the PML. For this example the calibration set-up is
identical to the problem of interest, including the evaluation time
$T_{e} = T_{c}$.

\begin{figure}
  \centering \subfloat[The solid blue curve shows the energy reduction
    for a constant attenuation function, and the dotted red curve for
    a piecewise-constant attenuation function as explained in
    Section~\ref{sec:pml}.
  \label{fig:ac_layerplot}]{
%
%
%
%
\begin{tikzpicture}

\begin{axis}[
xlabel={PML width [\centi\meter]},
ylabel={Energy reduction [$\deci\bel$]},
xmin=0, xmax=18,
ymin=-71, ymax=-10,
axis on top,
width=\figurewidth,
height=\figureheight,
xtick={0,2,4,6,8,10,12,14,16,18},
xticklabels={0,2,4,6,8,10,12,14,16,18},
ytick={-70,-60,-50,-40,-30,-20,-10},
yticklabels={-70,-60,-50,-40,-30,-20,-10},
legend entries={{Constant},{Piecewise-constant}}
]
\addplot [myblue, mark=*]
coordinates {
(0.999999999999999,-19.0679758601392)
(2,-30.2260613762608)
(3,-36.3341231650073)
(4,-41.087940896615)
(5,-45.2902206397836)
(6,-49.0867010966273)
(7,-51.9216029026512)
(8,-53.8167264516859)
(9,-54.9894818330264)
(10,-55.6713191655659)
(11,-56.4278102783693)
(12,-57.235071821526)
(13,-58.0405295762739)
(14,-58.8192043492471)
(15,-59.5576186311786)
(16,-60.2490929838224)
(17,-60.8917663265251)
(18,-61.4871155139049)

};
\addplot [myred, mark=*]
coordinates {
(0.999999999999999,-19.0679758601392)
(2,-37.748305587841)
(3,-51.4895411541104)
(4,-55.6600877290788)
(5,-58.4796271362394)
(6,-63.3209722047065)
(7,-67.2677078973166)
(8,-69.5108101875915)
(9,-69.6121608183089)
(10,-69.1696996078998)
(11,-69.2204973530631)
(12,-67.9279998092156)
(13,-67.3194048842274)
(14,-66.4384517120418)
(15,-66.9297138276899)
(16,-66.7003905759139)
(17,-66.7235637541637)
(18,-68.4355419485824)

};
\path [draw=black, fill opacity=0] (axis cs:-4.44089209850063e-16,-10)--(axis cs:18,-10);

\path [draw=black, fill opacity=0] (axis cs:18,-71)--(axis cs:18,-10);

\path [draw=black, fill opacity=0] (axis cs:-4.44089209850063e-16,-71)--(axis
cs:18,-70);

\path [draw=black, fill opacity=0] (axis 
cs:-4.44089209850063e-16,-71)--(axis cs:-4.44089209850063e-16,-10);

\path [dotted, draw=black, fill opacity=0] 
(axis cs:0,-45.2902206397836)--(axis cs:5,-45.2902206397836);

\path [dotted, draw=black, fill opacity=0] 
(axis cs:5,-45.2902206397836)--(axis cs:5,-71);

\end{axis}

\end{tikzpicture}}
  \hfil
  \subfloat[Perfectly matched layers are considered with three different
  widths~$w$.
  \label{fig:ac_polyplot}]{
%
%
%
%
\begin{tikzpicture}

\begin{axis}[
xlabel={Polynomial degree},
ylabel={Energy reduction [$\deci\bel$]},
xmin=0, xmax=6,
ymin=-71, ymax=-10,
axis on top,
width=\figurewidth,
height=\figureheight,
xtick={0,1, 2, 3, 4, 5, 6},
xticklabels={0, 1, 2, 3, 4, 5, 6},
ytick={-70,-60,-50,-40, -30},
yticklabels={-70,-60,-50,-40, -30},
legend entries={{$w$ = 0.05~\meter},{$w$ = 0.10~\meter},{$w$ = 0.15~\meter}}
]
\addplot [myblue, mark=*]
coordinates {
    (0,-45.5866141650945)
    (1,-43.8921634182227)
    (2,-43.0663986119135)
    (3,-42.8819038744709)
    (4,-44.3349840912732)
    (5,-42.3434833242494)
    (6,-42.3349992050573)
};
\addplot [myred, mark=*]
coordinates {
    (0,-55.6713191655659)
    (1,-55.3707108030877)
    (2,-59.208390149025)
    (3,-59.4528348309133)
    (4,-67.3203087458811)
    (5,-58.8940532433996)
    (6,-56.5038232434347)
};
\addplot [myblack, mark=*]
coordinates {
    (0,-59.5576186311786)
    (1,-60.176154101174)
    (2,-70.6747840775695)
    (3,-66.8360290758371)
    (4,-63.6168462206943)
    (5,-64.3424939366394)
    (6,-64.1302429387845)
};
\path [draw=black, fill opacity=0] (axis cs:0,-10)--(axis cs:20,-10);

\path [draw=black, fill opacity=0] (axis cs:20,-71)--(axis cs:20,-40);

\path [draw=black, fill opacity=0] (axis cs:0,-71)--(axis cs:20,-71);

\path [draw=black, fill opacity=0] (axis cs:0,-71)--(axis cs:0,-40);

\path [dotted, draw=black, fill opacity=0]
(axis cs:2,-44.2250703252688)--(axis cs:2,-70);

\end{axis}

\end{tikzpicture}}
  \caption{Energy reduction with perfectly matched layers for the
    acoustic wave example depending on the perfectly matched layer
    width (left) and polynomial degree for the attenuation function
    (right).}
\end{figure}
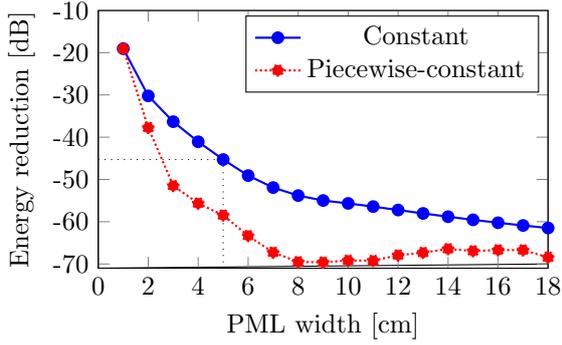
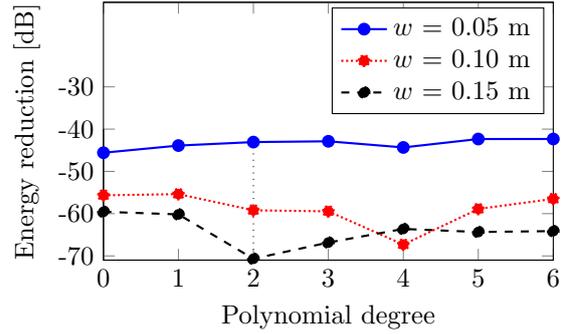

We first compare constant with piecewise-constant AFs for different
PML widths. The smallest considered PML is $0.01~\meter$ wide. The PML
is extended $0.01~\meter$ in $x$-direction seventeen times, up to a
total width of $0.18~\meter$. When a piecewise-constant AF is
considered, one control value is added for every extension, e.g., for
a $0.05~\meter$ wide PML, the piecewise-constant AF is defined by five
control variables. The energy reduction, as defined
in~\eqref{eq:energyreduction}, for these experiments with the
calibrated AFs is shown in Figure~\ref{fig:ac_layerplot}. These
results show that piecewise-constant AFs perform better than
constant~AFs for every PML width.

The reduction in energy for different polynomial order AFs and
different PML widths is shown in Figure~\ref{fig:ac_polyplot}.  First
note the results for the fourth-order polynomial AF, where the
$0.10~\meter$ PML appears to outperform the $0.15~\meter$ PML. This
peculiarity points to the optimisation problem being non-convex. We
will comment on this further when examining initial guesses for the
controls.  Comparing Figure~\ref{fig:ac_layerplot} and
Figure~\ref{fig:ac_polyplot}, it can be concluded that
piecewise-constant AFs outperform the polynomial~AFs, e.g., for a
$0.10~\meter$ wide PML, the calibrated piecewise-constant AF reduces
the energy more than any polynomial~AF. We note from
Figure~\ref{fig:ac_polyplot} that there appears to be limited benefit
in using polynomial orders greater than two, which is consistent with
finite difference results presented by \citet{chew96jin}. We restrict
further experiments to AFs to polynomial degrees of two or less.

We would expect the performance of a polynomial AF to be at least as
good as the constant AF case since the polynomial case contains the
constant case. However, Figure~\ref{fig:ac_polyplot} shows that for a
$0.05~\meter$ wide PML, a constant AF is slightly more effective than
any other polynomial~AF. This again points to the optimisation problem
being non-convex.

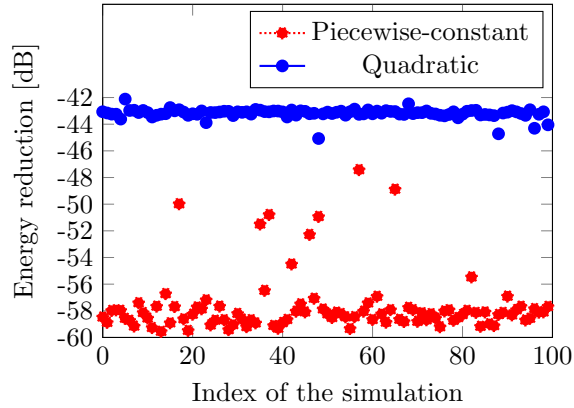
\begin{figure}
  \centering
%
%
%
%
\begin{tikzpicture}

\begin{axis}[
xlabel={Index of the simulation},
ylabel={Energy reduction [dB]},
xmin=0, xmax=100,
ymin=-60, ymax=-35,
axis on top,
width=\figurewidth,
height=1.2\figureheight,
xtick={0,20,40,60,80,100},
xticklabels={0,20,40,60,80,100},
ytick={-60,-58,-56,-54,-52,-50,-48,-46,-44,-42},
yticklabels={-60,-58,-56,-54,-52,-50,-48,-46,-44,-42},
legend entries={{Piecewise-constant},{Quadratic}}
]
\addplot [myred, draw=none, mark=*]
coordinates {
(0,-58.4796271362394)
(1,-58.8772794015011)
(2,-57.9875861003059)
(3,-57.9409226441235)
(4,-57.9682197486934)
(5,-58.5751354188154)
(6,-58.7160468639612)
(7,-59.154612432601)
(8,-57.4039355130995)
(9,-58.1433684376227)
(10,-58.5517841041867)
(11,-59.2690889349943)
(12,-57.6629423098273)
(13,-59.5539361588308)
(14,-56.7217499645163)
(15,-58.9120339894111)
(16,-57.6892812964382)
(17,-49.980445776599)
(18,-58.6106112584299)
(19,-59.4829071224565)
(20,-58.2673947654222)
(21,-57.7211926126882)
(22,-57.8600014593036)
(23,-57.1853736695165)
(24,-59.0682597690714)
(25,-58.7127484691479)
(26,-57.6574265746641)
(27,-58.6771228627827)
(28,-59.4210178034174)
(29,-59.0474801669615)
(30,-58.2054483969538)
(31,-58.6297779053526)
(32,-59.1998014707295)
(33,-58.6790526310825)
(34,-58.9149537235423)
(35,-51.4979230241081)
(36,-56.4757228747094)
(37,-50.7878480894231)
(38,-59.1125862536033)
(39,-59.3200910343248)
(40,-58.8961039358214)
(41,-58.6589679502092)
(42,-54.4959946952583)
(43,-58.04130850707)
(44,-57.4885703909971)
(45,-58.092559114369)
(46,-52.2796609813827)
(47,-57.0696464825886)
(48,-50.9317722264762)
(49,-57.8733615912242)
(50,-58.2394690685874)
(51,-58.5146606674973)
(52,-58.107158900666)
(53,-58.2132317716214)
(54,-58.4585512660512)
(55,-59.33902002789)
(56,-58.4492027227324)
(57,-47.4206264103199)
(58,-58.0414720703397)
(59,-57.4157865131413)
(60,-58.6592449594498)
(61,-56.9198574339538)
(62,-58.2422176420162)
(63,-58.8672097200786)
(64,-57.9234128501084)
(65,-48.8852047985097)
(66,-58.6459057512125)
(67,-58.8124297982861)
(68,-57.7491254243461)
(69,-57.9193671352418)
(70,-58.7992702791003)
(71,-58.1531416034985)
(72,-58.6839970685297)
(73,-58.2271813446953)
(74,-58.5521474011114)
(75,-59.2063206037366)
(76,-58.0230342710848)
(77,-57.9320320313388)
(78,-58.7211986792787)
(79,-58.5215672406783)
(80,-58.2874566290323)
(81,-57.974290025459)
(82,-55.4668536319108)
(83,-58.1306414680316)
(84,-59.1727869423121)
(85,-58.0945657312214)
(86,-58.9827237481735)
(87,-59.1193914304419)
(88,-58.2857396506494)
(89,-58.1322329897268)
(90,-56.9108995465832)
(91,-58.3168616855896)
(92,-57.9297345674475)
(93,-57.6684637003809)
(94,-58.6995764613891)
(95,-58.569129276332)
(96,-57.8476019314564)
(97,-58.1133984569354)
(98,-58.0924295371347)
(99,-57.680090823192)

};
\addplot [myblue, mark=*, draw=none]
coordinates {
    (0,-43.0663986119135)
    (1,-43.1598144721383)
    (2,-43.2442167501428)
    (3,-43.2505283707619)
    (4,-43.6160713805877)
    (5,-42.1171019735492)
    (6,-42.9710423698645)
    (7,-42.9534373659626)
    (8,-43.1263626108653)
    (9,-42.9691043312793)
    (10,-43.1287058643952)
    (11,-43.4605701090042)
    (12,-43.3305069228957)
    (13,-43.2468735227396)
    (14,-43.2134265502501)
    (15,-42.7518047707967)
    (16,-42.9967176088933)
    (17,-42.9134674234659)
    (18,-43.0897666809053)
    (19,-43.3349450002081)
    (20,-43.217909648201)
    (21,-43.2563351784829)
    (22,-43.0042485337803)
    (23,-43.8835643498342)
    (24,-43.1207507850089)
    (25,-43.1292154536783)
    (26,-43.0860995838683)
    (27,-43.0283236650456)
    (28,-43.0354684357194)
    (29,-43.3542413591663)
    (30,-43.0716449454429)
    (31,-43.1116695014076)
    (32,-43.0719804527347)
    (33,-43.2447577698904)
    (34,-42.8911368002084)
    (35,-42.9455974816589)
    (36,-43.0555990492867)
    (37,-43.0730131481463)
    (38,-43.0083332512997)
    (39,-43.0142953770348)
    (40,-43.0767640925092)
    (41,-43.4599036789541)
    (42,-42.9353808439611)
    (43,-43.3213058947719)
    (44,-42.9960576011479)
    (45,-43.0345197122127)
    (46,-43.2146440982509)
    (47,-43.1826962876631)
    (48,-45.0743360550559)
    (49,-43.2260628983309)
    (50,-43.0933599804692)
    (51,-43.1997170677789)
    (52,-43.1040216836417)
    (53,-43.0218870521073)
    (54,-43.3247093334754)
    (55,-43.063603516616)
    (56,-42.8686694809297)
    (57,-43.2609662381463)
    (58,-43.2259361206663)
    (59,-42.929722272093)
    (60,-43.2205691833496)
    (61,-43.0176547734645)
    (62,-43.1766100929772)
    (63,-42.9835107990443)
    (64,-43.0427824585779)
    (65,-43.051013227951)
    (66,-43.241489346092)
    (67,-43.2675355762587)
    (68,-42.4686710839225)
    (69,-43.2073385543973)
    (70,-43.132891656868)
    (71,-43.1335453945635)
    (72,-43.0070618941257)
    (73,-43.1866108715985)
    (74,-43.2339071489038)
    (75,-43.3308376843605)
    (76,-43.3738502061149)
    (77,-43.2715570952114)
    (78,-43.0700325959672)
    (79,-43.5211598437589)
    (80,-43.2507935484552)
    (81,-43.0291722318418)
    (82,-42.9734554213913)
    (83,-42.9573489707342)
    (84,-43.3082540132981)
    (85,-43.268792999104)
    (86,-43.3041909797936)
    (87,-43.3585383102327)
    (88,-44.7202170684922)
    (89,-43.1639838288491)
    (90,-43.0981324008592)
    (91,-42.9834302677018)
    (92,-43.0744738730774)
    (93,-43.1971191704344)
    (94,-43.3375530990769)
    (95,-42.9158330852622)
    (96,-44.3058072708652)
    (97,-43.2554048464825)
    (98,-43.0700886671444)
    (99,-44.0476065696697)
};
\path [draw=black, fill opacity=0] (axis cs:0,-35)--(axis cs:100,-35);

\path [draw=black, fill opacity=0] (axis cs:100,-60)--(axis cs:100,-42);

\path [draw=black, fill opacity=0] (axis cs:0,-60)--(axis cs:100,-60);

\path [draw=black, fill opacity=0] (axis cs:0,-60)--(axis cs:0,-42);

\end{axis}

\end{tikzpicture}
  \caption{Energy reduction achieved by the calibrated attenuation
    functions for the acoustic wave example with zero initial values
    (index 0) and random sets of initial values (index $>0$). The
    perfectly matched layer is $0.05~\meter$ wide. The experiment was
    performed for quadratic (dotted red) and piecewise-constant
    attenuation functions (solid blue).}
  \label{fig:ac_rands}
\end{figure}

We now fix the PML width to $0.05~\meter$ to examine the influence of
the initial AF parameters. Figure~\ref{fig:ac_rands} shows the
reduction in energy after optimisation for zero initial values (index
$0$) and random starting values (indices greater than zero) for both a
piecewise-constant and a quadratic AF.  The starting values are
uniformly sampled on the interval $[0, 7000]$ for the
piecewise-constant case and the interval $[-500, 500]$ for the
polynomial case. For the polynomial case, we allow negative
coefficients in order to allow AFs that are not monotonically
increasing. For the piecewise-constant AFs, the energy reduction for
approximately ten percent of the results is more than $10~\deci\bel$
from the best result.  There is less variation in the computed energy
reduction for quadratic AFs compared to the piecewise-constant
case. However, every piecewise-constant AF outperforms all quadratic
AFs. In the remainder we will set the initial guess for all controls
to zero.

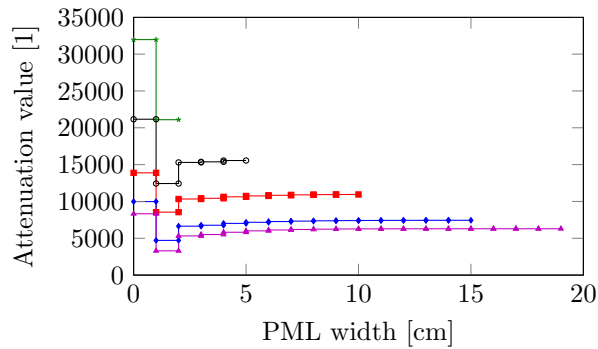
\begin{figure}
  \centering
%
%
%
%
\begin{tikzpicture}

\definecolor{color1}{rgb}{0.75,0,0.75}
\definecolor{color0}{rgb}{0,0.75,0.75}
\definecolor{color2}{rgb}{0.75,0.75,0}

\begin{axis}[
xlabel={PML width [\centi\meter]},
ylabel={Attenuation value [1]},
xmin=0, xmax=20,
ymin=0, ymax=35000,
axis on top,
scaled y ticks = false,
width=\figurewidth,
height=\figureheight,
xtick={0,5,10,15,20},
xticklabels={0,5,10,15,20},
ytick={0,5000,10000,15000,20000,25000,30000,35000},
yticklabels={0,5000,10000,15000,20000,25000,30000,35000}
]
\addplot [green!50.0!black, mark=star, mark size=1]
coordinates {
	(0,31970.1025789044)
	(1,31970.1025789044)
	(1,21095.9147292999)
	(2,21095.9147292999)
	
};

\addplot [black, mark=o, mark size=1]
coordinates {
	(0,21161.645565)
	(1,21161.645565)
	(1,12432.582045)
	(2,12432.582045)
	(2,15297.69577)
	(3,15297.69577)
	(3,15376.659807)
	(4,15376.659807)
	(4,15556.60067)
	(5,15556.60067)
	
};

\addplot [red, mark=square*, mark size=1]
coordinates {
(0,13883.799217)
(1,13883.799217)
(1,8546.8783675)
(2,8546.8783675)
(2,10332.20595)
(3,10332.20595)
(3,10428.882488)
(4,10428.882488)
(4,10630.196431)
(5,10630.196431)
(5,10748.254823)
(6,10748.254823)
(6,10835.404174)
(7,10835.404174)
(7,10893.947739)
(8,10893.947739)
(8,10929.754976)
(9,10929.754976)
(9,10946.604495)
(10,10946.604495)

};

\addplot [blue, mark=diamond*, mark size=1]
coordinates {
(0,9982.9006451)
(1,9982.9006451)
(1,4714.1793432)
(2,4714.1793432)
(2,6644.0492943)
(3,6644.0492943)
(3,6769.8671159)
(4,6769.8671159)
(4,7021.7896684)
(5,7021.7896684)
(5,7169.9479629)
(6,7169.9479629)
(6,7267.940897)
(7,7267.940897)
(7,7331.1254094)
(8,7331.1254094)
(8,7372.2333191)
(9,7372.2333191)
(9,7399.2431203)
(10,7399.2431203)
(10,7417.1503823)
(11,7417.1503823)
(11,7429.0183973)
(12,7429.0183973)
(12,7436.7070126)
(13,7436.7070126)
(13,7441.3245528)
(14,7441.3245528)
(14,7443.4933302)
(15,7443.4933302)

};

\addplot [color1, mark=triangle*, mark size=1]
coordinates {
(0,8323.7825551)
(1,8323.7825551)
(1,3295.3145387)
(2,3295.3145387)
(2,5317.5984122)
(3,5317.5984122)
(3,5507.3921098)
(4,5507.3921098)
(4,5817.9066217)
(5,5817.9066217)
(5,6008.6058583)
(6,6008.6058583)
(6,6128.3687815)
(7,6128.3687815)
(7,6199.2579867)
(8,6199.2579867)
(8,6239.7512407)
(9,6239.7512407)
(9,6262.340558)
(10,6262.340558)
(10,6274.8708407)
(11,6274.8708407)
(11,6281.9720769)
(12,6281.9720769)
(12,6286.2306622)
(13,6286.2306622)
(13,6289.0102996)
(14,6289.0102996)
(14,6290.9818193)
(15,6290.9818193)
(15,6292.4455677)
(16,6292.4455677)
(16,6293.5175833)
(17,6293.5175833)
(17,6294.2313341)
(18,6294.2313341)
(18,6294.5897923)
(19,6294.5897923)

};

\path [draw=black, fill opacity=0] (axis cs:0,35000)--(axis cs:20,35000);

\path [draw=black, fill opacity=0] (axis cs:20,0)--(axis cs:20,35000);

\path [draw=black, fill opacity=0] (axis cs:0,0)--(axis cs:20,0);

\path [draw=black, fill opacity=0] (axis cs:0,0)--(axis cs:0,35000);

\end{axis}

\end{tikzpicture}
  \caption{Optimal piecewise-constant attenuation functions for
    perfectly matched layers for the acoustic wave example for the
    case of $0.02~\meter$ (green stars), $0.05~\meter$ (black
    circles), $0.10~\meter$ (red squares), $0.15~\meter$ (blue
    diamonds) and $0.19~\meter$ (purple triangles) wide perfectly
    matched layer.}
  \label{fig:ac_pcplots}
\end{figure}

Figure~\ref{fig:ac_pcplots} shows the piecewise-constant AF for
$0.02~\meter$, $0.05~\meter$, $0.10~\meter$, $0.15~\meter$ and
$0.19~\meter$ wide PMLs. The result is not immediately intuitive; the
first control value is relatively large, followed by a substantially
smaller second control value. The remaining control values are
approximately equal and larger than the second value. The
counter-intuitive outcome highlights an advantage of using an
optimisation approach.

\begin{figure}
  \centering
  \begin{minipage}{0.45\textwidth}
%
%
%
%
\begin{tikzpicture}

\definecolor{color1}{rgb}{0.75,0,0.75}
\definecolor{color0}{rgb}{0,0.75,0.75}
\definecolor{color2}{rgb}{0.75,0.75,0}

\begin{groupplot}[group style={group size=1 by 2}]
\nextgroupplot[
ylabel={Value of the controls [1]},
xmin=0, xmax=30,
ymin=0, ymax=18000,
axis on top,
width=\figurewidth,
height=1.2\figureheight,
legend style={at={(0.03,0.97)}, anchor=north west},
legend
entries={{$c_0$},{$c_1$},{$c_2$},{$c_3$},{$c_4$},{$c_5$},{$c_6$},{$c_7$}}
]
\addplot [myblue]
coordinates {
(0,0)
(1,0.35349094029)
(2,1.7674547014)
(3,7.423309746)
(4,30.046729924)
(5,120.54041064)
(6,942.92570945)
(7,1476.1861247)
(8,2118.2180507)
(9,2718.6781005)
(10,3336.6918944)
(11,3950.3227381)
(12,4569.4214098)
(13,5192.3070292)
(14,5823.1094099)
(15,6465.9825027)
(16,7129.0567518)
(17,7826.082856)
(18,8582.1286011)
(19,9444.7774171)
(20,10510.071034)
(21,11980.746668)
(22,14195.976439)
(23,16092.399517)
(24,15786.268075)
(25,15642.850198)
(26,15531.921396)
(27,15693.619238)
(28,16021.185235)
(29,16239.677844)

};
\addplot [mygreen]
coordinates {
(0,0)
(1,0.35351037286)
(2,1.7675518643)
(3,7.4237178301)
(4,30.048381693)
(5,120.54703715)
(6,942.94268533)
(7,1476.0548743)
(8,2117.5121562)
(9,2716.7271057)
(10,3332.202236)
(11,3941.155387)
(12,4551.8375603)
(13,5159.9657816)
(14,5765.1572948)
(15,6363.9015049)
(16,6951.0219953)
(17,7516.7866527)
(18,8043.4954789)
(19,8497.7781234)
(20,8815.5449886)
(21,8874.492878)
(22,8481.7388308)
(23,7987.8579132)
(24,8224.8843561)
(25,8560.0475137)
(26,9486.095915)
(27,10041.556494)
(28,10534.550286)
(29,10925.845026)

};
\addplot [myred]
coordinates {
(0,0)
(1,0.35353063769)
(2,1.7676531884)
(3,7.4241433914)
(4,30.050104203)
(5,120.55394745)
(6,942.99419728)
(7,1476.1323525)
(8,2117.6255908)
(9,2716.9074633)
(10,3332.5603484)
(11,3941.9606845)
(12,4553.6928341)
(13,5164.1022452)
(14,5773.9920166)
(15,6382.0312574)
(16,6987.0766489)
(17,7586.9410347)
(18,8178.4203062)
(19,8756.978441)
(20,9318.0899025)
(21,9864.179935)
(22,10409.210683)
(23,10776.10437)
(24,10806.621441)
(25,10945.806902)
(26,11426.1753)
(27,11783.023158)
(28,12145.023597)
(29,12426.863801)

};
\addplot [mycyan]
coordinates {
(0,0)
(1,0.35355043313)
(2,1.7677521656)
(3,7.4245590957)
(4,30.051786816)
(5,120.5606977)
(6,943.03969826)
(7,1476.1756492)
(8,2117.6100149)
(9,2716.7652801)
(10,3332.2141726)
(11,3941.3661085)
(12,4552.8818214)
(13,5163.2760896)
(14,5773.6657947)
(15,6383.2656603)
(16,6991.8330301)
(17,7598.5916707)
(18,8202.3916789)
(19,8801.2342729)
(20,9392.1845705)
(21,9972.5014007)
(22,10527.101219)
(23,10844.086241)
(24,10877.367463)
(25,11005.309653)
(26,11439.697728)
(27,11777.708689)
(28,12159.661919)
(29,12457.019389)

};
\addplot [mypurple]
coordinates {
(0,0)
(1,0.35356839654)
(2,1.7678419827)
(3,7.4249363273)
(4,30.053313706)
(5,120.56682322)
(6,943.08096959)
(7,1476.215715)
(8,2117.6020569)
(9,2716.659182)
(10,3331.9647732)
(11,3940.9746268)
(12,4552.4394582)
(13,5163.0435535)
(14,5774.1919153)
(15,6385.5698806)
(16,6997.6894757)
(17,7610.9726868)
(18,8226.176558)
(19,8844.352121)
(20,9467.4448208)
(21,10100.220241)
(22,10734.233022)
(23,11106.973486)
(24,11129.455293)
(25,11248.247671)
(26,11663.967571)
(27,11997.306501)
(28,12377.471117)
(29,12671.393501)

};
\addplot [myyellow]
coordinates {
(0,0)
(1,0.35358322475)
(2,1.7679161238)
(3,7.4252477198)
(4,30.054574104)
(5,120.57187964)
(6,943.11473291)
(7,1476.2472445)
(8,2117.5912118)
(9,2716.5648621)
(10,3331.7520486)
(11,3940.6482873)
(12,4552.0749318)
(13,5162.839773)
(14,5774.5430794)
(15,6387.1673663)
(16,7001.6693246)
(17,7619.1295425)
(18,8241.2812111)
(19,8870.6186065)
(20,9511.178744)
(21,10170.464289)
(22,10840.937042)
(23,11235.911034)
(24,11254.973736)
(25,11371.460645)
(26,11783.659794)
(27,12117.629551)
(28,12498.954747)
(29,12792.907884)

};
\addplot [myblack]
coordinates {
(0,0)
(1,0.35359379996)
(2,1.7679689998)
(3,7.4254697991)
(4,30.055472996)
(5,120.57548579)
(6,943.13868486)
(7,1476.2691318)
(8,2117.5820333)
(9,2716.4959959)
(10,3331.6007449)
(11,3940.4213236)
(12,4551.8286546)
(13,5162.712458)
(14,5774.7957463)
(15,6388.238041)
(16,7004.2474702)
(17,7624.2589987)
(18,8250.5081391)
(19,8886.2047864)
(20,9536.3983417)
(21,10209.919549)
(22,10899.747424)
(23,11307.108507)
(24,11324.933395)
(25,11441.514436)
(26,11856.044164)
(27,12193.220168)
(28,12577.816126)
(29,12873.859454)

};
\addplot [myorange]
coordinates {
(0,0)
(1,0.35359930379)
(2,1.767996519)
(3,7.4255853796)
(4,30.055940822)
(5,120.57736259)
(6,943.15111014)
(7,1476.2803303)
(8,2117.5767799)
(9,2716.4595785)
(10,3331.5219211)
(11,3940.3045219)
(12,4551.7037465)
(13,5162.6499655)
(14,5774.9250135)
(15,6388.7701853)
(16,7005.5054878)
(17,7626.7176136)
(18,8254.8490768)
(19,8893.3936973)
(20,9547.7876886)
(21,10227.348797)
(22,10925.185382)
(23,11337.650914)
(24,11355.127362)
(25,11472.062045)
(26,11888.519762)
(27,12227.675161)
(28,12614.277713)
(29,12911.712104)

};
\path [draw=black, fill opacity=0] (axis cs:13,18000)--(axis cs:13,18000);

\path [draw=black, fill opacity=0] (axis cs:30,13)--(axis cs:30,13);

\path [draw=black, fill opacity=0] (axis cs:13,0)--(axis cs:13,0);

\path [draw=black, fill opacity=0] (axis cs:0,13)--(axis cs:0,13);

\path [draw=black, dashed] (axis cs:16,7005)--(axis cs:16,0);

\nextgroupplot[
xlabel style={align=center},
xlabel={Index of the optimization step},
ylabel={Energy reduction $[\deci\bel]$},
yshift=-3mm,
xmin=0, xmax=30,
ymin=-70, ymax=0,
axis on top,
width=\figurewidth,
height=.75\figureheight,
]
\addplot [myblue]
coordinates {
(0,0)
(1,-0.00174202399621548)
(2,-0.0087101202238475)
(3,-0.036582502620372)
(4,-0.14807198745674)
(5,-0.59402892203233)
(6,-4.64631596206311)
(7,-7.27254185602863)
(8,-10.4313397868808)
(9,-13.3806494940744)
(10,-16.4083878763456)
(11,-19.4027475742162)
(12,-22.4054244778988)
(13,-25.3979436854344)
(14,-28.3837177991544)
(15,-31.3557925688737)
(16,-34.3086527209282)
(17,-37.2337124048855)
(18,-40.1246233689848)
(19,-42.9932527040248)
(20,-45.9381255757098)
(21,-49.4315844756603)
(22,-55.5898805822754)
(23,-60.1397231850751)
(24,-60.8210264886351)
(25,-61.6384307327364)
(26,-63.9280184304144)
(27,-65.9186615042806)
(28,-68.3677695374643)
(29,-69.5108101875915)

};
\path [draw=black, fill opacity=0] (axis cs:13,0)--(axis cs:13,0);

\path [draw=black, fill opacity=0] (axis cs:30,13)--(axis cs:30,13);

\path [draw=black, fill opacity=0] (axis cs:13,-70)--(axis cs:13,-70);

\path [draw=black, fill opacity=0] (axis cs:0,13)--(axis cs:0,13);

\path [draw=black, dashed] (axis cs:16,0)--(axis cs:16,-70);

\end{groupplot}

\end{tikzpicture}
  \end{minipage}
  \caption{Evolution of controls (top) and reduction in energy
    (bottom) for the acoustic wave example as a function of the
    iteration step during the optimisation process for calibrating a
    piecewise-constant attenuation function with eight controls on a
    $0.08~\meter$ wide perfectly matched layer.}
  \label{fig:ac_8layers}
\end{figure}
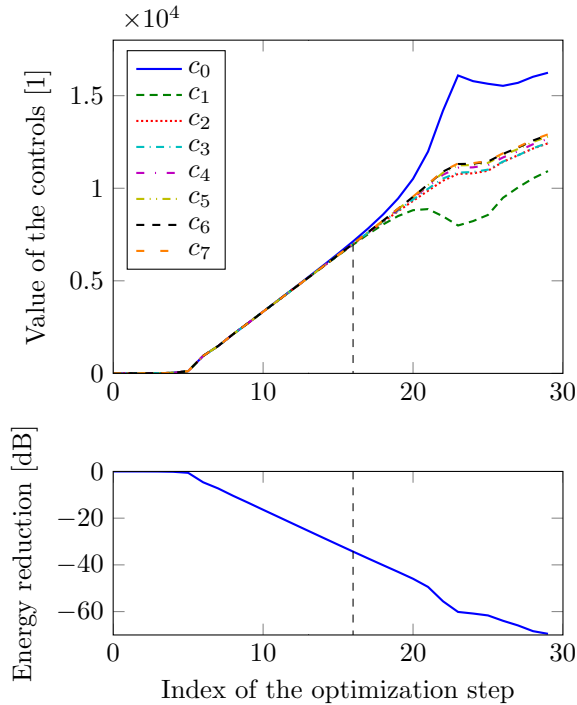

Figure~\ref{fig:ac_8layers} shows how the control values of the
piecewise-constant AF change with each optimiser iteration for a
$0.08~\meter$ wide PML together with the corresponding reduction in
energy. For up to approximately 17 iterations the process favours a
constant~AF. From the point at which the AF deviates significantly
from a constant AF, a further $20\deci\bel$ to $30\deci\bel$ reduction
in energy is observed.

\subsubsection{Elastodynamic wave propagation on a square}
\label{sec:elastoresultspml}

\begin{figure}
  \centering
  \subfloat[\label{fig:elastodynamic}]{
    \includegraphics[width=.45\textwidth]{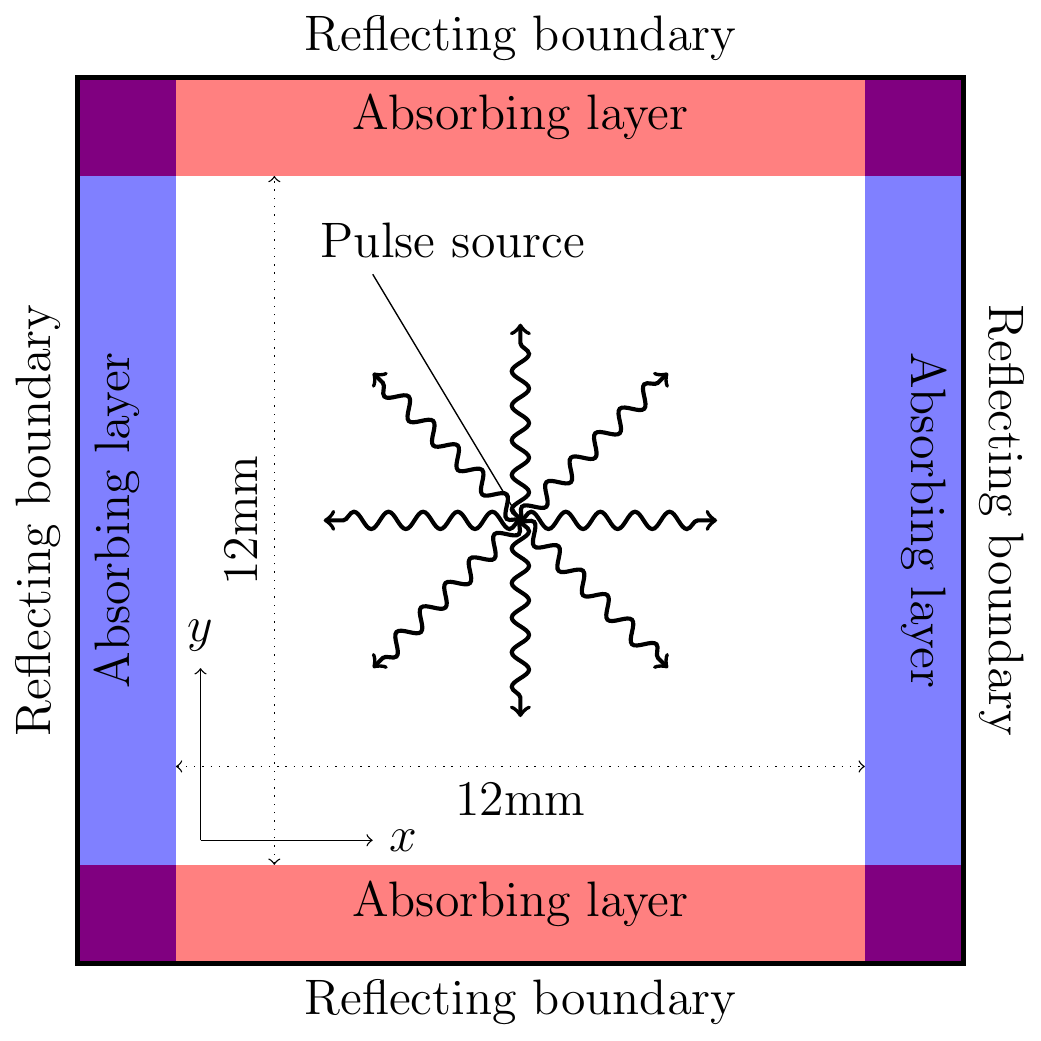}}
  \hfil
  \subfloat[\label{fig:elastodynamicfront}]{
    \includegraphics[width=.45\textwidth]{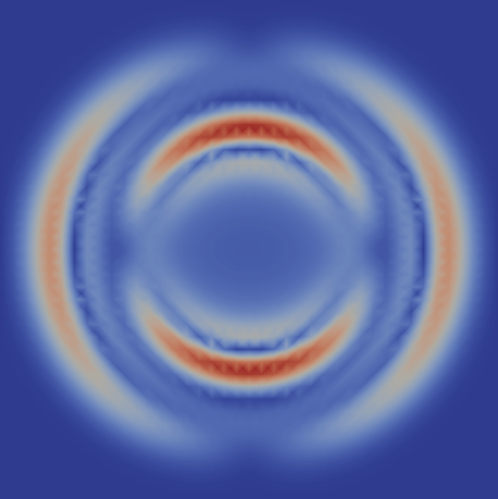}}
  \caption{Geometry of the elastodynamic wave example with waves
    propagating in radial direction (left) with typical resulting wave
    propagation pattern (right).}
\end{figure}

We simulate elastic wave propagation in an isotropic, homogeneous
square domain $\Omega_{i} = [-6, 6]^{2}~\milli\meter$, which is
extended in both the $x$- and $y$-directions with a $6~\milli\meter$
wide PML (see Figure~\ref{fig:elastodynamic}). We implement reflecting
fixed boundaries on all sides of the computational domain. The
longitudinal wave speed in the considered medium is $v_{l} =
5830.95~\meter\per \second$ and the transverse wave speed~$v_{t} =
3464.10~\meter\per \second$. The mass density of the considered
material is $2500~\kilo\gram\per\cubic\meter$. A Gaussian source
$\vec{f} = (f_{x}, 0)$ is applied, where
\begin{equation}
  f_{x} =
  \exp\del{-\del{\frac{t - 50 \Delta t}{50 \Delta t/4}}^{2}}
  \exp\del{-\del{\frac{x}{10^{-6}}}^{2}}
  \exp\del{-\del{\frac{y}{10^{-6}}}^{2}}.
\end{equation}
A typical resulting elliptical wave front for this example is
illustrated in Figure~\ref{fig:elastodynamicfront}.

The domain is meshed with crossed-triangle cells with edge length
$1.2~\milli\meter$. We solve this example using a discontinuous
Galerkin finite element method, which is presented
in~\ref{sec:eddiscr}. A time step size of $\Delta t = 4 \times
10^{-8}~\second$ is used. The calibration time is chosen to be $T_{c}
= t_0 + (12~\milli\meter)\sqrt{2}/c_t$, which is the time needed for
the peak to enter the domain and travel to the corner of the
computational domain at the lowest wave speed.
We consider the problem of interest to be identical to the calibration set-up
with the exception of the evaluation time. Since the calibration time doesn't
allow the wave to travel once in both direction through the absorbing layer,
the reduction in energy would not include the full benefit of the absorbing
layer. The evaluation time for this example is $T_{e} = t_0 +
(24~\milli\meter)\sqrt{2}/c_{t}$, which is the calibration time plus
the time needed for the peak to travel back from the corner of the
computational domain to the centre of the domain.

This model uses one AF in each spatial direction. It is however
undesirable to have orientation dependent PMLs because of the symmetry
of the domain. We therefore choose to define the AFs in both
directions by the same control variables.

The evolution of the controls and reduction in energy at each
optimiser step for the piecewise-constant case are shown in
Figure~\ref{fig:ed_layerhistorybound}. Despite the large changes in
the control values at low iteration counts, the reduction in energy
remains more-or-less constant from the second iteration. The final
result is again not a monotonically increasing function, shown by the
AF in Figure~\ref{fig:ed_shapes} (solid blue line).
\begin{figure}
  \centering
  \begin{minipage}{0.45\textwidth}
%
%
%
%
\begin{tikzpicture}

\definecolor{color1}{rgb}{0.75,0,0.75}
\definecolor{color0}{rgb}{0,0.75,0.75}

\begin{groupplot}[group style={group size=1 by 2}]
\nextgroupplot[
ylabel={Value of the controls [1]},
xmin=0, xmax=25,
ymin=0, ymax=250,
axis on top,
width=\figurewidth,
height=\figureheight,
xtick={0,5,10,15,20,25},
xticklabels={0,5,10,15,20,25},
ytick={0,50,100,150,200,250},
yticklabels={0,50,100,150,200,250},
legend entries={{$c_0$},{$c_1$},{$c_2$},{$c_3$},{$c_4$}},
legend columns=2
]
\addplot [myblue]
coordinates {
(0,0)
(1,228.35073966)
(2,76.016557323)
(3,25.314611604)
(4,8.4393100151)
(5,8.4393117902)
(6,8.4393101469)
(7,8.4393102469)
(8,8.4393101475)
(9,8.4393101469)
(10,8.439311922)
(11,8.4393102623)
(12,8.4393102635)
(13,8.4393102623)
(14,9.7979896273)
(15,12.086335723)
(16,14.126140844)
(17,16.208042955)
(18,17.931986551)
(19,19.128527294)
(20,19.700887243)
(21,19.8514476)

};
\addplot [mygreen]
coordinates {
(0,0)
(1,235.74373798)
(2,78.477641012)
(3,26.134188022)
(4,8.7125379665)
(5,8.7125379665)
(6,8.7125379665)
(7,8.7125379665)
(8,8.7125379665)
(9,8.7125379665)
(10,8.7125379665)
(11,8.7125379665)
(12,8.7125379665)
(13,8.7125379665)
(14,8.7125700589)
(15,8.7127184307)
(16,8.7129339292)
(17,8.7132348904)
(18,8.7135797945)
(19,8.7139529516)
(20,8.7143180147)
(21,8.7146470466)

};
\addplot [myred]
coordinates {
(0,0)
(1,230.78566799)
(2,76.827129994)
(3,25.584544012)
(4,8.5292992794)
(5,8.5292992794)
(6,8.5292992794)
(7,8.5292992794)
(8,8.5292992794)
(9,8.5292992794)
(10,8.5292992794)
(11,8.5292992794)
(12,8.5292992794)
(13,8.5292992794)
(14,8.5292828464)
(15,8.5292248159)
(16,8.5291463)
(17,8.5290400909)
(18,8.5289213476)
(19,8.52879588)
(20,8.52867582)
(21,8.5285692497)

};
\addplot [mycyan]
coordinates {
(0,0)
(1,201.03898844)
(2,66.924643256)
(3,22.286872893)
(4,7.4299314779)
(5,7.4299314779)
(6,7.4299314779)
(7,7.4299314779)
(8,7.4299314779)
(9,7.4299314779)
(10,7.4299314779)
(11,7.4299314779)
(12,7.4299314779)
(13,7.4299314779)
(14,7.4299171699)
(15,7.429866629)
(16,7.4297982408)
(17,7.4297057287)
(18,7.4296022973)
(19,7.4294930079)
(20,7.429388428)
(21,7.4292955983)

};
\addplot [mypurple]
coordinates {
(0,0)
(1,152.43300733)
(2,50.744010975)
(3,16.898488623)
(4,5.6335679372)
(5,5.6335679372)
(6,5.6335679372)
(7,5.6335679372)
(8,5.6335679372)
(9,5.6335679372)
(10,5.6335679372)
(11,5.6335679372)
(12,5.6335679372)
(13,5.6335679372)
(14,5.6335570886)
(15,5.6335187674)
(16,5.6334669138)
(17,5.6333967688)
(18,5.6333183445)
(19,5.6332354785)
(20,5.6331561834)
(21,5.6330857975)

};
\path [draw=black, fill opacity=0] (axis cs:0,250)--(axis cs:25,250);

\path [draw=black, fill opacity=0] (axis cs:25,0)--(axis cs:25,250);

\path [draw=black, fill opacity=0] (axis cs:0,0)--(axis cs:25,0);

\path [draw=black, fill opacity=0] (axis cs:0,0)--(axis cs:0,250);

\nextgroupplot[
xlabel={Index of the optimization step},
ylabel={Energy reduction $[dB]$},
xmin=0, xmax=25,
ymin=-48.85, ymax=-48.3,
axis on top,
width=\figurewidth,
height=\figureheight,
xtick={0,5,10,15,20,25},
xticklabels={0,5,10,15,20,25},
ytick={-48.9,-48.8,-48.7,-48.6,-48.5,-48.4,-48.3},
yticklabels={-48.85,-48.80,-48.75,-48.70,-48.65,$\ldots$,0.}
]

\addplot [mydblue]
coordinates {
(0,-48.3)
(1,-48.8149131198133)

};
\addplot [myblue]
coordinates {
(1,-48.8149131198133)
(2,-48.7988171706414)
(3,-48.7362747658228)
(4,-48.4842058550083)
(5,-48.4847807747942)
(6,-48.4840416638477)
(7,-48.4842307835028)
(8,-48.4843648345923)
(9,-48.4840416638477)
(10,-48.4838237581382)
(11,-48.4841641849423)
(12,-48.4845131437509)
(13,-48.4841641849423)
(14,-48.5406197319902)
(15,-48.6049920694287)
(16,-48.6435360059675)
(17,-48.6719225086847)
(18,-48.6898833007575)
(19,-48.7001379732486)
(20,-48.7046722320055)
(21,-48.7057617940677)

};
\path [draw=black, fill opacity=0] (axis cs:0,-48.3)--(axis cs:25,-48.3);

\path [draw=black, fill opacity=0] (axis cs:25,-48.85)--(axis cs:25,-48.45);

\path [draw=black, fill opacity=0] (axis cs:0,-48.85)--(axis cs:25,-48.85);

\path [draw=black, fill opacity=0] (axis cs:0,-48.85)--(axis cs:0,-48.45);

\end{groupplot}

\end{tikzpicture}
  \end{minipage}
  \caption{Evolution of the controls (top) and reduction in energy
    (bottom) for the elastodynamic wave example as a function of the
    iteration step during the calibration process for
    a piecewise-constant attenuation function with five
    parameters for a $6~\milli\meter$ wide perfectly matched layer.}
    \label{fig:ed_layerhistorybound}
\end{figure}
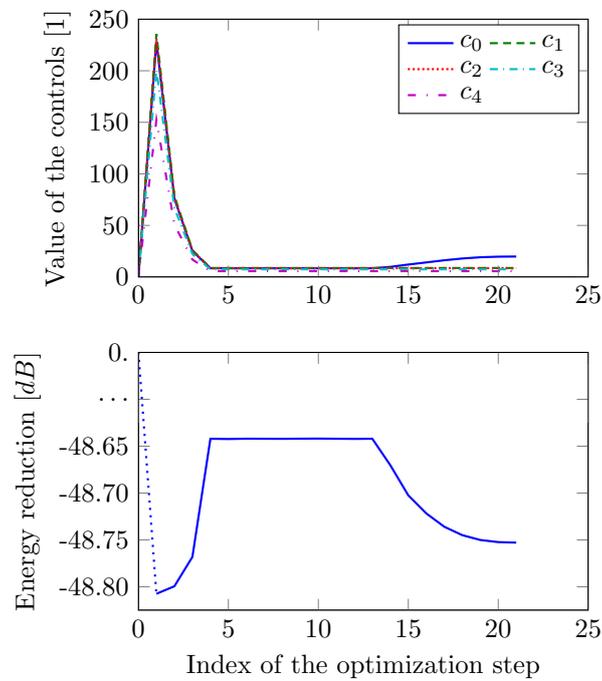
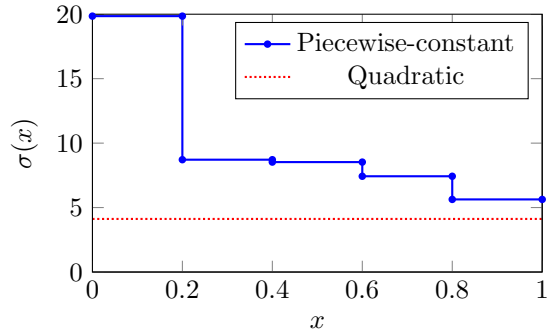
\begin{figure}
  \centering
  \begin{minipage}{0.45\textwidth}
%
%
%
%
\begin{tikzpicture}

\begin{axis}[
xmin=0, xmax=1,
ymin=0, ymax=20,
axis on top,
width=\figurewidth,
height=\figureheight,
xlabel={$x$},
ylabel={$\sigma(x)$},
ytick={0,5, 10 ,15, 20},
yticklabels={0,5, 10 ,15, 20},
legend entries={{Piecewise-constant},{Quadratic }},
legend columns=1,
legend style={at={(0.97,0.97)}, anchor=north east}
]
\addplot [myblue, mark=*, mark size=1]
coordinates {
(0,19.8514476)
(0.2,19.8514476)
(0.2,8.7146470466)
(0.4,8.7146470466)
(0.4,8.5285692497)
(0.6,8.5285692497)
(0.6,7.4292955983)
(0.8,7.4292955983)
(0.8,5.6330857975)
(1,5.6330857975)

};
\addplot [myred]
coordinates {
(0,4.118)
(1,4.118)

};
\path [draw=black, fill opacity=0] (axis cs:0,600)--(axis cs:1,600);

\path [draw=black, fill opacity=0] (axis cs:1,0)--(axis cs:1,600);

\path [draw=black, fill opacity=0] (axis cs:0,0)--(axis cs:1,0);

\path [draw=black, fill opacity=0] (axis cs:0,0)--(axis cs:0,600);

\end{axis}

\end{tikzpicture}
  \end{minipage}
  \caption{Piecewise-constant (solid blue) and quadratic (dotted red)
    attenuation function $\sigma(x)$ obtained with the calibration procedure
    for the elastodynamic wave example with square geometry.}
  \label{fig:ed_shapes}
\end{figure}

\begin{figure}
  \centering
  \begin{minipage}{0.45\textwidth}
%
%
%
\begin{tikzpicture}

\begin{groupplot}[group style={group size=1 by 2}]
\nextgroupplot[
ylabel={Value of the control [1]},
xmin=0, xmax=70,
ymin=0, ymax=1200,
axis on top,
width=\figurewidth,
height=\figureheight,
xtick={0,10,20,30,40,50,60,70},
xticklabels={0,10,20,30,40,50,60,70},
ytick={0,200,400,600,800,1000,1200},
yticklabels={0,200,400,600,800,1000,1200},
legend entries={{$c_0$},{$c_1$},{$c_2$}}
]
\addplot [ultra thick, myblue]
coordinates {
(0,0)
(1,1029.9074776)
(2,342.79812964)
(3,114.10497826)
(4,37.98819046)
(5,12.653963115)
(6,12.653962083)
(7,12.65396309)
(8,12.653963115)
(9,12.653963115)
(10,12.653963115)
(11,12.653963115)
(12,12.653963115)
(13,12.653963115)
(14,12.653963115)
(15,12.653963115)
(16,12.653963115)
(17,12.653963115)
(18,12.653963115)
(19,12.653963115)
(20,12.653962083)
(21,12.653963084)
(22,12.653963115)
(23,12.653963115)
(24,12.653963099)
(25,12.653963115)
(26,12.653963115)
(27,12.653963115)
(28,12.653963115)
(29,12.653963115)
(30,12.653833266)
(31,12.65331387)
(32,12.651236286)
(33,12.642925952)
(34,12.609684615)
(35,12.476719268)
(36,11.94485788)
(37,9.8174123259)
(38,1.3076301106)
(39,6.3180877537)
(40,3.5452464835)
(41,4.3727852034)
(42,3.5452464835)
(43,8.1488007418)
(44,4.0096264141)
(45,3.8610239535)
(46,3.8292462127)
(47,3.814377039)
(48,3.7715049305)
(49,3.7373687007)
(50,3.7338507698)
(51,3.9029037658)
(52,4.0558553363)
(53,4.1225733702)
(54,4.132254059)
(55,4.1420226058)
(56,4.1606691551)
(57,4.1901909013)
(58,4.2416399084)
(59,4.2190148985)
(60,4.1114931826)
(61,4.1190805497)
(62,4.1186740178)

};
\addplot [ultra thick, mygreen]
coordinates {
(0,0)
(1,502.61848135)
(2,167.29335311)
(3,55.685847645)
(4,18.539108622)
(5,6.175424358)
(6,6.1754243069)
(7,6.1754243568)
(8,6.175424358)
(9,6.175424358)
(10,6.175424358)
(11,6.175424358)
(12,6.175424358)
(13,6.175424358)
(14,6.175424358)
(15,6.175424358)
(16,6.175424358)
(17,6.175424358)
(18,6.175424358)
(19,6.175424358)
(20,6.1754243069)
(21,6.1754243565)
(22,6.175424358)
(23,6.175424358)
(24,6.1754243572)
(25,6.175424358)
(26,6.175424358)
(27,6.175424358)
(28,6.175424358)
(29,6.175424358)
(30,6.1754188216)
(31,6.1753966761)
(32,6.1753080942)
(33,6.1749537664)
(34,6.173536455)
(35,6.1678672097)
(36,6.1451902284)
(37,6.0544823032)
(38,5.6916506024)
(39,5.9052815535)
(40,5.7870558823)
(41,5.822339662)
(42,5.7870558823)
(43,5.8338042223)
(44,5.7917715845)
(45,5.7695573169)
(46,5.7391801448)
(47,5.6825780268)
(48,5.3012695786)
(49,4.382464793)
(50,1.3363871963)
(51,0)
(52,0)
(53,0)
(54,0)
(55,0)
(56,0)
(57,0)
(58,0)
(59,0)
(60,0)
(61,0)
(62,0)

};
\addplot [ultra thick, myred]
coordinates {
(0,0)
(1,316.4100798)
(2,105.31507529)
(3,35.055542426)
(4,11.670802122)
(5,3.887573948)
(6,3.8875739448)
(7,3.887573948)
(8,3.887573948)
(9,3.887573948)
(10,3.887573948)
(11,3.887573948)
(12,3.887573948)
(13,3.887573948)
(14,3.887573948)
(15,3.887573948)
(16,3.887573948)
(17,3.887573948)
(18,3.887573948)
(19,3.887573948)
(20,3.8875739448)
(21,3.8875739479)
(22,3.887573948)
(23,3.887573948)
(24,3.887573948)
(25,3.887573948)
(26,3.887573948)
(27,3.887573948)
(28,3.887573948)
(29,3.887573948)
(30,3.8875736116)
(31,3.8875722657)
(32,3.8875668821)
(33,3.887545348)
(34,3.8874592113)
(35,3.8871146648)
(36,3.8857364787)
(37,3.8802237342)
(38,3.858172756)
(39,3.8711561073)
(40,3.8639709808)
(41,3.866115341)
(42,3.8639709808)
(43,3.8652975526)
(44,3.8641047977)
(45,3.8626314469)
(46,3.8605470599)
(47,3.8566452213)
(48,3.8303376552)
(49,3.7669218758)
(50,3.5566478955)
(51,2.9217588959)
(52,2.380039915)
(53,2.1276670263)
(54,2.0742091897)
(55,1.987894429)
(56,1.7315956713)
(57,1.0989195572)
(58,0)
(59,0)
(60,0)
(61,0)
(62,0)

};
%
%
%

\nextgroupplot[
xlabel={Index of the optimization step},
ylabel={Energy reduction $[dB]$},
xmin=0, xmax=70,
ymin=-46.2, ymax=-45.4,
axis on top,
width=\figurewidth,
height=\figureheight,
xtick={0,10,20,30,40,50,60, 70},
xticklabels={0,10,20,30,40,50,60,70},
ytick={-46.2,-46.1,-46.0,-45.9,-45.8,-45.7,-45.6,-45.5,-45.4},
yticklabels={-46.2,-46.1,-46.0,-45.9,-45.8,-45.7,-45.6,$\ldots$,0},
]

\addplot [ultra thick, blue, dotted]
coordinates {
(0,-45.8)
(1,-46.0899908203443)

};
\addplot [ultra thick, blue]
coordinates {
(1,-46.0899908203443)
(2,-46.0768714461086)
(3,-46.0451426685112)
(4,-46.0023770069232)
(5,-46.0628223592296)
(6,-46.0628226147885)
(7,-46.0628212124516)
(8,-46.0628223592296)
(9,-46.0628223592296)
(10,-46.0628223592296)
(11,-46.0628223592296)
(12,-46.0628223592296)
(13,-46.0628223592296)
(14,-46.0628223592296)
(15,-46.0628223592296)
(16,-46.0628223592296)
(17,-46.0628223592296)
(18,-46.0628223592296)
(19,-46.0628223592296)
(20,-46.0628226147885)
(21,-46.0628210987759)
(22,-46.0628223592296)
(23,-46.0628223592296)
(24,-46.0628209084139)
(25,-46.0628223592296)
(26,-46.0628223592296)
(27,-46.0628223592296)
(28,-46.0628223592296)
(29,-46.0628223592296)
(30,-46.0628238981366)
(31,-46.062828932446)
(32,-46.0628506879743)
(33,-46.0629387553147)
(34,-46.0632928632869)
(35,-46.0647254544432)
(36,-46.0706790392768)
(37,-46.0978792524727)
(38,-45.6309560931425)
(39,-46.1385209089933)
(40,-46.0659200472101)
(41,-46.1136266628018)
(42,-46.0659200472101)
(43,-46.1209237504742)
(44,-46.0968835569696)
(45,-46.0882104578735)
(46,-46.0860903276928)
(47,-46.084930371028)
(48,-46.0805412690084)
(49,-46.0741881323162)
(50,-46.059461426501)
(51,-46.0664828056448)
(52,-46.077582929496)
(53,-46.0819923058928)
(54,-46.0826058646498)
(55,-46.0832109525924)
(56,-46.0843226571629)
(57,-46.0859810466009)
(58,-46.0887795189398)
(59,-46.0874226860736)
(60,-46.0806036444833)
(61,-46.0811051030813)
(62,-46.081078911668)

};
%
%
%

\end{groupplot}

\end{tikzpicture}
  \end{minipage}
  \caption{Evolution of the controls (top) and reduction in energy
    (bottom) for the elastodynamic wave example with square geometry
    as a function of the iteration step during the optimisation
    process for calibrating a quadratic attenuation function for a
    $6~\milli\meter$ wide perfectly matched layer.}
  \label{fig:ed_polyhistory_clean}
\end{figure}
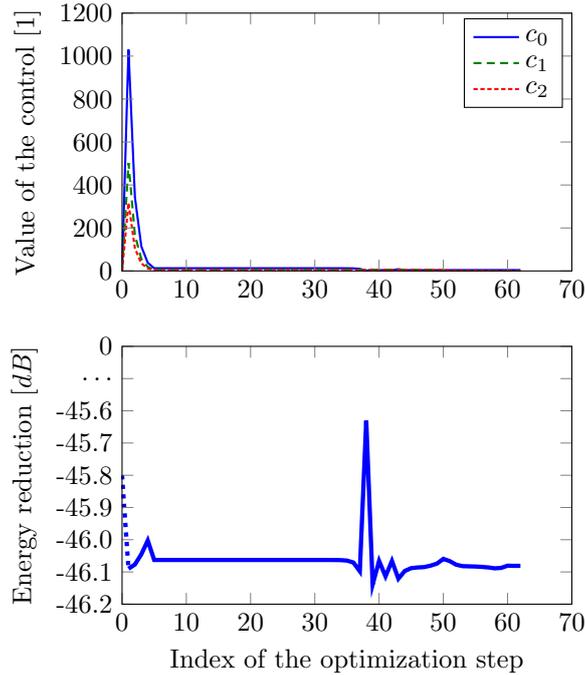

As a second experiment for this model, we compare the
piecewise-constant result to a quadratic AF. The evolution of the
controls and reduction in energy during the optimisation process for a
quadratic AF are shown in Figure~\ref{fig:ed_polyhistory_clean}.  The
calibration process for a quadratic AF has resulted in a constant AF
(see Figure~\ref{fig:ed_shapes}) and performs about $5\%$ less well
than the calibrated piecewise-constant AF.

\subsection{Consecutive matched layers}
\label{sec:cmlresults}

We now move to examining the performance of the truncation strategy of
consecutive matched layers presented in Section~\ref{sec:cml}. In this
case each `sub-layer' has a constant attenuation function associated
with it. The key difference with perfectly matched layers is the
absence of auxiliary fields and equations in the model.

\subsubsection{Elastodynamic wave propagation on a square}
\label{sec:cmlsquare}

\begin{figure}
  \centering
  \begin{minipage}{0.45\textwidth}
%
%
%
%
\begin{tikzpicture}

\definecolor{color1}{rgb}{0.75,0,0.75}
\definecolor{color0}{rgb}{0,0.75,0.75}

\begin{groupplot}[group style={group size=1 by 2}]
\nextgroupplot[
ylabel={Value of the controls [1]},
xmin=0, xmax=25,
ymin=0, ymax=250,
axis on top,
width=\figurewidth,
height=\figureheight,
xtick={0,5,10,15,20,25},
xticklabels={0,5,10,15,20,25},
ytick={0,50,100,150,200,250},
yticklabels={0,50,100,150,200,250},
legend entries={{$c_0$},{$c_1$},{$c_2$},{$c_3$},{$c_4$}},
legend columns=2
]
\addplot [myblue]
coordinates {
(0,0)
(1,215.81129716)
(2,71.843451334)
(3,23.926105053)
(4,7.9776139262)
(5,7.977616004)
(6,7.9776243153)
(7,7.9776575603)
(8,7.9777905407)
(9,7.9783224619)
(10,7.9804501469)
(11,7.9889608868)
(12,8.0230038464)
(13,8.159175685)
(14,8.7038630392)
(15,11.182542734)
(16,13.100324472)
(17,15.288469916)
(18,17.170128105)
(19,18.643048515)
(20,19.489480968)
(21,19.794562341)
(22,19.84437663)

};
\addplot [mygreen]
coordinates {
(0,0)
(1,223.65363442)
(2,74.454160703)
(3,24.795552518)
(4,8.2675113494)
(5,8.2675113493)
(6,8.2675113492)
(7,8.2675113485)
(8,8.2675113457)
(9,8.2675113344)
(10,8.2675112894)
(11,8.2675111095)
(12,8.2675103899)
(13,8.2675075114)
(14,8.2674959975)
(15,8.2675480188)
(16,8.2676805726)
(17,8.2679176981)
(18,8.2682133435)
(19,8.2685669714)
(20,8.2689438652)
(21,8.2693100873)
(22,8.2696369854)

};
\addplot [myred]
coordinates {
(0,0)
(1,219.40044081)
(2,73.038275103)
(3,24.324018551)
(4,8.1102891048)
(5,8.1102891048)
(6,8.1102891048)
(7,8.1102891048)
(8,8.1102891048)
(9,8.1102891048)
(10,8.1102891049)
(11,8.1102891051)
(12,8.1102891061)
(13,8.1102891102)
(14,8.1102891265)
(15,8.1102562819)
(16,8.1102017703)
(17,8.1101124973)
(18,8.1100068075)
(19,8.1098855474)
(20,8.1097611056)
(21,8.109643617)
(22,8.109540219)

};
\addplot [mycyan]
coordinates {
(0,0)
(1,195.2605407)
(2,65.002116839)
(3,21.647727765)
(4,7.2179409939)
(5,7.2179409939)
(6,7.2179409939)
(7,7.2179409939)
(8,7.2179409939)
(9,7.2179409939)
(10,7.217940994)
(11,7.2179409942)
(12,7.2179409952)
(13,7.2179409992)
(14,7.217941015)
(15,7.2179117881)
(16,7.2178632746)
(17,7.2177838232)
(18,7.2176897615)
(19,7.2175818443)
(20,7.2174710978)
(21,7.217366541)
(22,7.2172745247)

};
\addplot [mypurple]
coordinates {
(0,0)
(1,151.02001092)
(2,50.274471019)
(3,16.742963385)
(4,5.582559199)
(5,5.582559199)
(6,5.582559199)
(7,5.582559199)
(8,5.582559199)
(9,5.582559199)
(10,5.582559199)
(11,5.5825591992)
(12,5.5825592)
(13,5.5825592031)
(14,5.5825592154)
(15,5.582536611)
(16,5.5824990896)
(17,5.58243764)
(18,5.5823648903)
(19,5.5822814244)
(20,5.5821957701)
(21,5.582114903)
(22,5.582043735)

};
\path [draw=black, fill opacity=0] (axis cs:0,250)--(axis cs:25,250);

\path [draw=black, fill opacity=0] (axis cs:25,0)--(axis cs:25,250);

\path [draw=black, fill opacity=0] (axis cs:0,0)--(axis cs:25,0);

\path [draw=black, fill opacity=0] (axis cs:0,0)--(axis cs:0,250);

\nextgroupplot[
xlabel={Index of the optimization step},
ylabel={Energy reduction $[dB]$},
xmin=0, xmax=25,
ymin=-48.85, ymax=-48.3,
axis on top,
width=\figurewidth,
height=\figureheight,
xtick={0,5,10,15,20,25},
xticklabels={0,5,10,15,20,25},
ytick={-48.9,-48.8,-48.7,-48.6,-48.5,-48.4,-48.3},
yticklabels={-48.85,-48.80,-48.75,-48.70,-48.65,$\ldots$,0.}
]

\addplot [mydblue]
coordinates {
(0,-48.3)
(1,-48.8144970205459)

};
\addplot [myblue]
coordinates {
(1,-48.8144970205459)
(2,-48.7972957295975)
(3,-48.7300049125192)
(4,-48.4602527883407)
(5,-48.4602529014573)
(6,-48.4602533539289)
(7,-48.4602551637943)
(8,-48.4602624031227)
(9,-48.4602913579313)
(10,-48.4604071374576)
(11,-48.4608696209073)
(12,-48.4627094544205)
(13,-48.4699105765556)
(14,-48.496384050928)
(15,-48.5828034397643)
(16,-48.625757806539)
(17,-48.6604272268775)
(18,-48.68242818101)
(19,-48.6961819358086)
(20,-48.7030231068339)
(21,-48.7053257320533)
(22,-48.7056940340232)

};
\path [draw=black, fill opacity=0] (axis cs:0,-48.3)--(axis cs:25,-48.3);

\path [draw=black, fill opacity=0] (axis cs:25,-48.85)--(axis cs:25,-48.45);

\path [draw=black, fill opacity=0] (axis cs:0,-48.85)--(axis cs:25,-48.85);

\path [draw=black, fill opacity=0] (axis cs:0,-48.85)--(axis cs:0,-48.45);

\end{groupplot}

\end{tikzpicture}
  \end{minipage}
  \caption{Evolution of the controls (top) and reduction in energy
    (bottom) for the elastodynamic wave example as a function of the
    iteration step during the bounded calibration process for
    consecutive matched layers.}
    \label{fig:cml_history}
\end{figure}
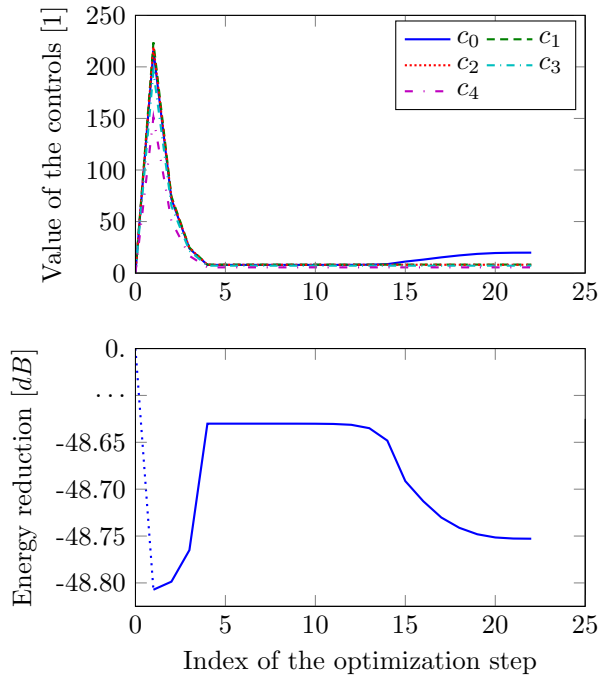

We revisit the elastodynamic example from
Section~\ref{sec:elastoresultspml}. Both the problem of interest and the
calibration set-up are identical to the previous example with the PML
replaced by CMLs. The evolution of the controls and reduction in energy
during the optimisation process for calibrating the five CMLs are shown in
Figure~\ref{fig:cml_history}. The results shown in
Figure~\ref{fig:cml_history} are almost identical to the PML results
in Figure~\ref{fig:ed_layerhistorybound}. Since the model for CMLs
does not require ADEs, in contrast to the PML model, the automatic
calibration procedure for CMLs is faster than for PMLs.

\subsubsection{Elastodynamic wave propagation on a more complicated
geometry}
\label{sec:arbitrary}

\begin{figure}
  \centering
  \includegraphics[width=.4\textwidth]{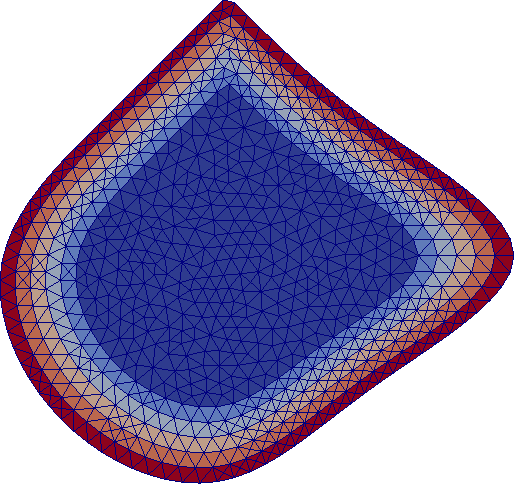}
  \caption{The more complicated computational domain used for the
    elastodynamic wave experiment with consecutive matched layers. The
    domain of interest (dark blue) is surrounded by five tightly
    wrapped consecutive matched layers, each indicated by a distinct
    colour.}
  \label{fig:arbitrary_domain}
\end{figure}

We now adopt the problem of interest and calibration set-up of the preceding
elastic example, but replace the square domain by the domain and mesh shown in
Figure~\ref{fig:arbitrary_domain}.  The domain of interest is shown in
dark blue. Five consecutive matched layers are placed around the
domain of interest. The precise definition of the domain and the mesh
are available in the supporting material~\citep{pmlcode}.

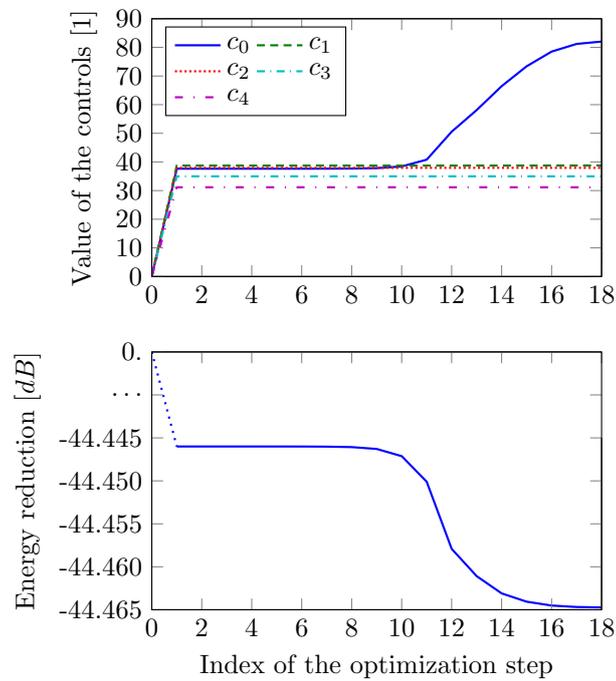
\begin{figure}
  \centering
  \begin{minipage}{0.45\textwidth}
%
%
%
%
\begin{tikzpicture}

\definecolor{color1}{rgb}{0.75,0,0.75}
\definecolor{color0}{rgb}{0,0.75,0.75}

\begin{groupplot}[group style={group size=1 by 2}]
\nextgroupplot[
ylabel={Value of the controls [1]},
xmin=0, xmax=18,
ymin=0, ymax=90,
axis on top,
width=\figurewidth,
height=\figureheight,
xtick={0,2,4,6,8,10,12,14,16,18},
xticklabels={0,2,4,6,8,10,12,14,16,18},
ytick={0,10,20,30,40,50,60,70,80,90},
yticklabels={0,10,20,30,40,50,60,70,80,90},
legend style={at={(0.03,0.97)}, anchor=north west},
legend entries={{$c_0$},{$c_1$},{$c_2$},{$c_3$},{$c_4$}},
legend columns=2
]
\addplot [myblue]
coordinates {
(-4.44089209850063e-16,0)
(0.999999999999999,37.634208127)
(2,37.63421716)
(3,37.63425329)
(4,37.63439781)
(5,37.634975892)
(6,37.637288221)
(7,37.646537534)
(8,37.683534786)
(9,37.831523795)
(10,38.423479829)
(11,40.791303968)
(12,50.552755447)
(13,58.098103)
(14,66.442923044)
(15,73.385418872)
(16,78.515021869)
(17,81.197126317)
(18,82.023057514)

};
\addplot [mygreen]
coordinates {
(-4.44089209850063e-16,0)
(0.999999999999999,38.731399187)
(2,38.731399189)
(3,38.7313992)
(4,38.731399243)
(5,38.731399414)
(6,38.7314001)
(7,38.731402843)
(8,38.731413814)
(9,38.731457698)
(10,38.731633235)
(11,38.732335384)
(12,38.735530165)
(13,38.738488835)
(14,38.7427123)
(15,38.747856686)
(16,38.754356197)
(17,38.761761099)
(18,38.769241432)

};
\addplot [myred]
coordinates {
(-4.44089209850063e-16,0)
(0.999999999999999,37.936947182)
(2,37.936947182)
(3,37.936947182)
(4,37.936947182)
(5,37.936947182)
(6,37.936947182)
(7,37.936947183)
(8,37.936947186)
(9,37.936947199)
(10,37.936947249)
(11,37.93694745)
(12,37.936861056)
(13,37.936652113)
(14,37.936144566)
(15,37.935248411)
(16,37.933802041)
(17,37.931881408)
(18,37.929778729)

};
\addplot [mycyan]
coordinates {
(-4.44089209850063e-16,0)
(0.999999999999999,34.98429124)
(2,34.98429124)
(3,34.98429124)
(4,34.98429124)
(5,34.98429124)
(6,34.98429124)
(7,34.98429124)
(8,34.984291241)
(9,34.984291245)
(10,34.984291262)
(11,34.984291327)
(12,34.984211084)
(13,34.984017838)
(14,34.983548941)
(15,34.982721436)
(16,34.981386188)
(17,34.979613338)
(18,34.977672562)

};
\addplot [mypurple]
coordinates {
(-4.44089209850063e-16,0)
(0.999999999999999,31.160679632)
(2,31.160679632)
(3,31.160679632)
(4,31.160679632)
(5,31.160679632)
(6,31.160679632)
(7,31.160679632)
(8,31.160679633)
(9,31.160679637)
(10,31.160679651)
(11,31.160679709)
(12,31.160608235)
(13,31.160436109)
(14,31.160018459)
(15,31.159281395)
(16,31.158092082)
(17,31.156512995)
(18,31.154784336)

};
\path [draw=black, fill opacity=0] (axis cs:-4.44089209850063e-16,90)--(axis cs:18,90);

\path [draw=black, fill opacity=0] (axis cs:18,0)--(axis cs:18,90);

\path [draw=black, fill opacity=0] (axis cs:-4.44089209850063e-16,0)--(axis cs:18,0);

\path [draw=black, fill opacity=0] (axis cs:-4.44089209850063e-16,0)--(axis cs:-4.44089209850063e-16,90);

\nextgroupplot[
xlabel={Index of the optimization step},
ylabel={Energy reduction $[dB]$},
xmin=0, xmax=18,
ymin=-44.465, ymax=-44.435,
axis on top,
width=\figurewidth,
height=\figureheight,
xtick={0,2,4,6,8,10,12,14,16,18},
xticklabels={0,2,4,6,8,10,12,14,16,18},
ytick={-44.47,-44.465,-44.46,-44.455,-44.45,-44.445,-44.44,-44.435},
yticklabels={,-44.465,-44.460,-44.455,-44.450,-44.445,$\ldots$,0.}
]

\addplot [mydblue]
coordinates {
(-4.44089209850063e-16,-44.435)
(0.999999999999999,-44.4460018451581)

};
\addplot [myblue]
coordinates {
(0.999999999999999,-44.4460018451581)
(2,-44.4460018584462)
(3,-44.4460019115954)
(4,-44.4460021241907)
(5,-44.4460029745495)
(6,-44.446006375579)
(7,-44.4460199731795)
(8,-44.446074259535)
(9,-44.4462897506075)
(10,-44.4471258957557)
(11,-44.4500955135359)
(12,-44.4579124216715)
(13,-44.4610854014395)
(14,-44.463093610496)
(15,-44.4640586521479)
(16,-44.4645094309001)
(17,-44.4646802077526)
(18,-44.4647252840937)

};
\path [draw=black, fill opacity=0] (axis cs:-4.44089209850063e-16,-44.435)--(axis cs:18,-44.435);

\path [draw=black, fill opacity=0] (axis cs:18,-44.465)--(axis cs:18,-44.445);

\path [draw=black, fill opacity=0] (axis cs:-4.44089209850063e-16,-44.465)--(axis cs:18,-44.465);

\path [draw=black, fill opacity=0] (axis cs:-4.44089209850063e-16,-44.465)--(axis cs:-4.44089209850063e-16,-44.445);

\end{groupplot}

\end{tikzpicture}
  \end{minipage}
  \caption{Evolution of the controls (top) and reduction in energy
    (bottom) for the elastodynamic wave example with a complicated
    geometry as a function of the iteration step during the
    calibration process for consecutive matched layers.}
  \label{fig:arb_history}
\end{figure}

The evolution of the attenuation constants for this problem and the
corresponding reduction in energy are shown in
Figure~\ref{fig:arb_history}. The reduction in energy is only ten
percent less than for the problem on the square domain. The
attenuation in layers closer to the domain of interest is larger than
for the layers farther from the domain of interest. This is probably a
manifestation of the sensitivity of the different controls. It is to
be expected that the attenuation in the outer layers has less effect
on the reduction in energy, since a considerable amount of energy will
have been damped by layers closer to the domain of interest.

\subsubsection{Electromagnetic wave propagation}
\label{sec:em}

We consider an application for which an absorbing region is
calibrated, and then used to solve a problem of interest.
The problem of interest involves a transverse electromagnetic
wave~\citep{jackson98} in a parallel plate wave guide. We solve
equation~\eqref{eq:emsyst} on the domain of interest~$\Omega_i = [0,
  L_{x}] \times [0, L_{y}] \times [0, L_{z}(x)]$. We consider
conducting plates at $x = 0$ and $x = L_x$, which are both modelled by
implementing perfect electric conducting boundary conditions
($\vec{n}\times \vec{E} = \vec{0}$) at $x = 0$ and~$x = L_x$.  The face at
$z = 0$ is a port through which waves are inserted into the wave
guide.  We consider the case where the plates are infinite in
$y$-direction, which is modelled by applying perfect magnetic
conducting boundary conditions ($\vec{n}\times \vec{H} = \vec{0}$) at~$y =
0$ and~$y = L_y$.  For $z \geq L_{z}(x)$ there is open space, which
will be modelled using CMLs.

\begin{figure}
  \centering
  \begin{tikzpicture}
    \node[anchor=south west,inner sep=0] at (0,0)
	 {\includegraphics[width=\textwidth]{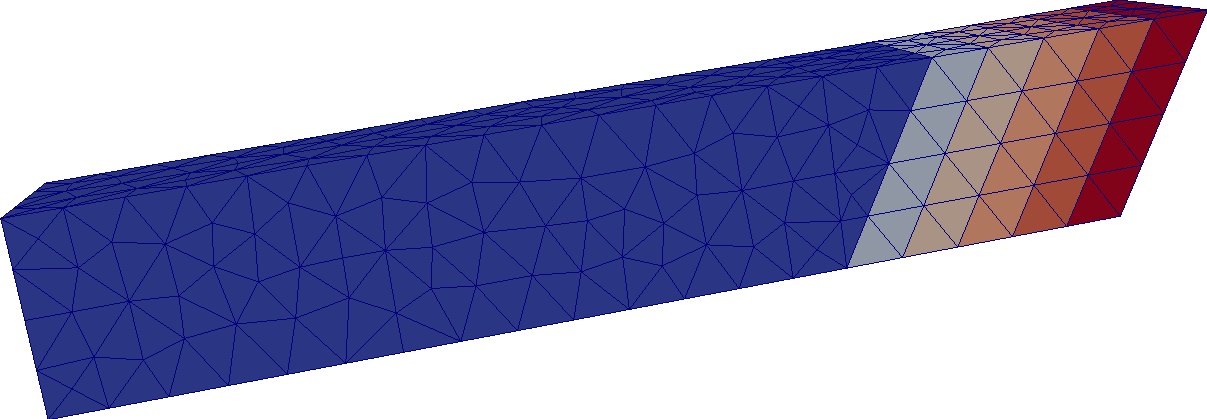}};
	 \draw[red,ultra thick,->] (10,3.4) -- (14,4.12);
	 \draw[black,ultra thick] (11.5,2.05) -- (12.7,4.9);
	 \draw[green,ultra thick] (11.3,3.6) arc (0:56:-1cm) node [midway, below,
	   green, anchor=west] {\Large $\theta$};
  \end{tikzpicture}
  \caption{A  parallel plate wave guide used for the
    electromagnetic example. The domain of interest (dark blue) is
    extended with five consecutive matched layers, each indicated by a
    different colour. In this problem, the waves enter the consecutive matched
    layers at a $60$~degree angle $\theta$. By modifying the length of
    the upper plate, the incidence angle can be controlled.}
  \label{fig:em_domain60}
\end{figure}

Before solving the problem of interest we calibrate the AFs on
the adsorbing layer. To study the impact of oblique incidence angles
at the boundary of the domain of interest, we
will `stretch' the upper conducting plate ($x = L_{x}$) in the $z$
direction. Three configurations will be tested, i.e., with $90$, $60$
and $45$~degree incidence angles. The domain with a $60$~degree
incidence angle is shown in Figure~\ref{fig:em_domain60}. The volume
of absorbing layers will slightly differ in all three cases due to the
different plate lengths, but the thickness of each layer (in the
$z$-direction) is fixed.

We extend the domain of interest with five absorbing layers, each one
cell wide (see Figure~\ref{fig:em_domain60}). Also the boundary
conditions of the domain of interest at $x=0$, $x=L_x$, $y=0$ and
$y=L_y$ are extended to the absorbing domain. The boundary condition
at the port ($z = 0 $) is set to $\vec{E} = \del{E_x, 0, 0}$, where
\begin{equation}
  E_x(t) = \exp\del{-\del{\frac{t - 10^{-8}}{10^{-8}/4}}^2}.
\end{equation}
At the end of the CMLs ($z = L_{z}(x)$) a perfect electric conducting
boundary condition is applied. The calibration time is chosen to be
the time for the peak of the input pulse to enter the system, move
through the domain of interest, reflect off the interface between the
domain of interest and the absorbing domain, and back to the source of
the input signal, which is $T_{c} = 10^{-8} + 2 L_{z}(0)/c$. To
optimise the attenuation functions, we initialise the AFs to zero, and
run the optimisation process for the $90$, $60$ and $45$ degree
incidence angle cases.

The evolution of the control variables and the corresponding reduction
in energy at $T_{e}=T_{c}$ for the cases with $90$, $60$ and $45$
degree incidence angles are shown in
Figure~\ref{fig:em_historyall}. We see that the two cases with
non-perpendicular incidence perform well relative to the to the $90$
degree case. The obtained attenuation values differ significantly
between the three cases. The smaller the incidence angle, the more
iterations are required to converge the optimisation algorithm.

We observe the least energy reduction for the $60$ degree incidence
case. The observation that the $45$ degree incidence case performs
better than both other cases is mainly because when a wave hits the
interface between the domain of interest and the absorbing domain at a
$45$ degree incidence angle, the wave is reflected to the upper plate,
hits it perpendicularly and hence is reflected again at a forty-five
degree angle to the CMLs, before it gets reflected again in negative
$z$-direction towards the source of the input signal. In other words,
reflected waves meet the CMLs for a second time sooner than in the
other cases.

\begin{figure}
  \centering
  \input{figures/em_all.tex}
  \caption{Evolution of the controls (left) and reduction in energy
    (right) for the electromagnetic wave example as a function of the
    iteration step during the calibration process for the consecutive
    matched layers for the domain with (a)~$90$ degree, (b)~$60$
    degree and (c)~$45$ degree incidence angles.}
  \label{fig:em_historyall}
\end{figure}

To complete the electromagnetic wave case study, we compute a
transverse electromagnetic wave in the wave guide with the $60$ degree
incidence angle and the CMLs that were calibrated for this case.  The
boundary conditions are as described for the calibration set-up,
except now as an input wave we apply the boundary condition $\vec{E} =
\del{E_x, 0, 0}$ at the port ($z = 0$), with
\begin{equation}
  E_x(t) = \sin\del{3.1 \times 10^{8} t}.
\end{equation}
There is no analytical solution available for a transverse
electromagnetic wave in a parallel plate wave guide where one plate is
longer than the other.  However, as the waves move from left to right
in the wave guide, the solution in the rectangular part $\mathcal{R} =
[0, L_{x}] \times [0, L_{y}] \times [0, L_{z}(0)]$ is not affected by
the rest of the domain. Hence, we can use the analytical solution for
a parallel wave guide with equal plates for $\mathcal{R}$ which is
\begin{equation}
  \vec{E} = \del{\sin\del{3.1 \times 10^{8} z/c - 3.1 \times 10^{8} t}, 0,
  0}, \qquad
  \vec{H} = \del{0, \frac{1}{c\mu_0} \sin\del{3.1 \times 10^{8} z/c - 3.1
  \times 10^{8}
      t}, 0, 0},
  \label{eq:emreference}
\end{equation}
where $c$ is the speed of light and $\mu_0$ is the permeability of
vacuum.

\begin{figure}
  \centering
  \input{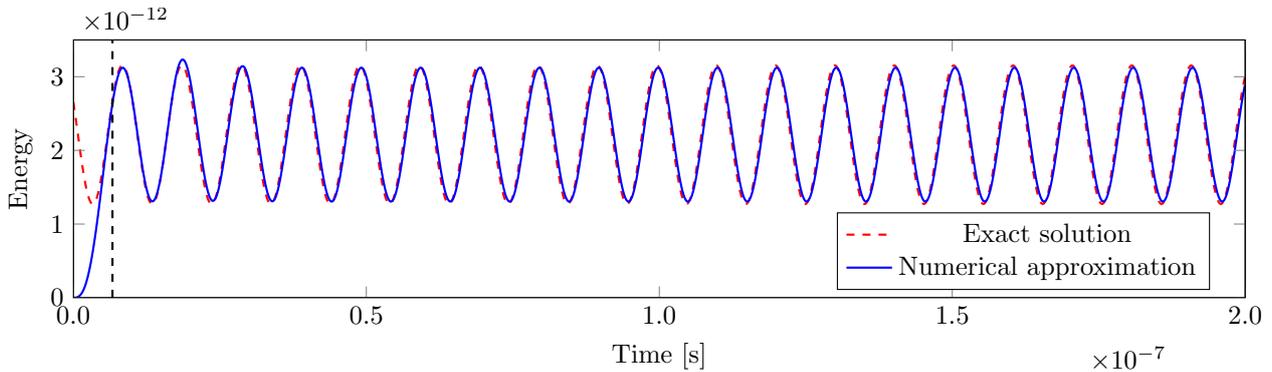}
  \caption{The computed electromagnetic energy (solid blue), as
    defined in~\eqref{eq:electromagneticenegery} in a parallel plate
    wave guide with output port under a 60 degree angle compared to
    the theoretical reference (dashed red). The reference is only
    valid once a steady state has been reached.}
  \label{fig:sine60}
\end{figure}

For the initial condition, it is not straightforward to extend the
analytical solution~\eqref{eq:emreference} into the absorbing region.
Therefore, we start with a zero initial value and compare the
numerical solution to the analytical solution after the problem
reaches a steady state.  To evaluate the numerical solution, we
compare the electromagnetic energy (defined
in~\eqref{eq:electromagneticenegery}) of the numerical solution
computed with third-order polynomial elements to the reference
solution in~\eqref{eq:emreference} in $\mathcal{R}$
Figure~\ref{fig:sine60}.  We see that, after reaching the steady
state, the periods of the numerical and exact solutions are well
aligned. Importantly, we see that there is no systematic increase in
energy for the numerical case, which demonstrates that the CMLs are
effective.

\section{Conclusions}
\label{sec:conclusions}

We have presented an approach to automatically calibrate attenuation
functions for matched layers in wave propagation problems solved using
finite element time domain methods. The presented procedure is not
problem-specific, and in principle can be used to calibrate perfectly
matched layers for any problem, regardless of the discretisation
method. We have experimentally shown that there is no need to use
polynomial attenuation functions higher than order
two. Piecewise-constant attenuation functions can however result in
equally effective perfectly matched layers. For piecewise-constant
attenuation functions, the calibration procedure does not prefer
monotonically increasing attenuation functions.

We have presented calibration of a damping strategy which we call
\emph{consecutive matched layers}. The automatic calibration procedure
for consecutive matched layers is identical to the calibration
procedure for perfectly matched layers. Consecutive matched layers
lead to a simpler model than perfectly matched layers, resulting in
shorter simulation times, for both the forward problem and the
calibration procedure. It was shown for a collection of examples that
consecutive matched layers can perform as well as perfectly matched
layers. As a major advantage of consecutive matched layers over
perfectly matched layers is that consecutive matched layers can be
easily applied to complex domains.

\section*{Acknowledgements}

The authors would like to acknowledge Febe Brackx for her help with
the preliminary implementation and simulations, Patrick Farrell and
Simon Funke for their assistance with dolfin-adjoint, Stefan
Vandewalle, for helpful discussions on the content of this paper.
Steven Vandekerckhove was been funded by a PhD grant from the Agency
for Innovation by Science and Technology (IWT) and an international
collaboration grant from the Research Foundation - Flanders (FWO).
\section*{References}
\bibliographystyle{plainnat}
\bibliography{bibtex/Articles,bibtex/Books,bibtex/Miscs}
\appendix
\section{Problem specific expressions}
\label{sec:specifics}

\subsection{Acoustic wave propagation}

The matrices $\vec{A}_{i}$ for the acoustic wave problem read:
\begin{equation}
\vec{A}_1 =
\begin{pmatrix}
0 & 0 & 0 & \rho^{-1} \\
0 & 0 & 0 & 0 \\
0 & 0 & 0 & 0 \\
K & 0 & 0 & 0  \\
\end{pmatrix}
, \quad
\vec{A}_2
=
\begin{pmatrix}
0 & 0 & 0 & 0 \\
0 & 0 & 0 & \rho^{-1} \\
0 & 0 & 0 & 0 \\
0 & K & 0 & 0  \\
\end{pmatrix}
,\quad
\vec{A}_3
=
\begin{pmatrix}
0 & 0 & 0 & 0 \\
0 & 0 & 0 & 0 \\
0 & 0 & 0 & \rho^{-1}\\
0 & 0 & K & 0  \\
\end{pmatrix}.
\label{eq:Aacoustic}
\end{equation}
The energy functional for the acoustic wave problem reads:
\begin{equation}
E_\text{acoustic}(t) = \frac{1}{2}\int_{\Omega} \rho \vec{v} \cdot
\vec{v}
+ \frac{1}{K} p^{2} \dif x
\label{eq:acousticenergy}
\end{equation}
The matrix $Q$ in~\eqref{eq:generalenergy} for this problem reads:
\begin{equation}
  Q = \text{diag}\del{\rho, \ \rho, \ \rho, \ \frac{1}{K}}.
\label{eq:Qacoustic}
\end{equation}

\subsection{Electromagnetic wave propagation}

The matrices $\vec{A}_{i}$ for the electromagnetic wave problem read:
\begin{multline}
  \vec{A}_1
  =
  \begin{pmatrix}
    0 & 0 & 0 & 0 & 0 & 0 \\
    0 & 0 & 0 & 0 & 0 & -\mu^{-1}\\
    0 & 0 & 0 & 0 & \mu^{-1} & 0\\
    0 & 0 & 0 & 0 & 0 & 0 \\
    0 & 0 & \varepsilon^{-1} & 0 & 0 & 0 \\
    0 & -\varepsilon^{-1} & 0 & 0 & 0 & 0 \\
  \end{pmatrix},
  \\
  \vec{A}_2
  =
  \begin{pmatrix}
    0 & 0 & 0 & 0 & 0 & \mu^{-1} \\
    0 & 0 & 0 & 0 & 0 & 0\\
    0 & 0 & 0 & -\mu^{-1} & 0 & 0\\
    0 & 0 & -\varepsilon^{-1} & 0 & 0 & 0 \\
    0 & 0 & 0 & 0 & 0 & 0 \\
    \varepsilon^{-1} & 0 & 0 & 0 & 0 & 0 \\
  \end{pmatrix},
  \\
  \vec{A}_3
  =
  \begin{pmatrix}
    0 & 0 & 0 & 0 & -\mu^{-1} & 0 \\
    0 & 0 & 0 & \mu^{-1} & 0 & 0\\
    0 & 0 & 0 & 0 & 0 & 0\\
    0 & \varepsilon^{-1} & 0 & 0 & 0 & 0 \\
    -\varepsilon^{-1} & 0 & 0 & 0 & 0 & 0 \\
    0 & 0 & 0 & 0 & 0 & 0 \\
  \end{pmatrix}.
  \label{eq:Aelectro}
\end{multline}
The energy functional for the electromagnetic wave problems reads:
\begin{equation}
  E_\text{em}(t) = \frac{1}{2}\int_{\Omega} \mu \vec{H} \cdot \vec{H} +
  \varepsilon \vec{E} \cdot \vec{E} \dif x.
  \label{eq:electromagneticenegery}
\end{equation}
The matrix $Q$ in~\eqref{eq:generalenergy} for this problem reads:
\begin{equation}
  Q = \text{diag}\del{\mu, \ \mu, \ \mu, \ \varepsilon,  \
    \varepsilon, \ \varepsilon}.
  \label{eq:Qelectro}
\end{equation}

\subsection{Elastodynamic wave propagation}

The matrices $\vec{A}_{i}$ for the elastodynamic wave problem read:
\begin{multline}
  \vec{A}_1
  =
  \begin{pmatrix}
    0 & 0 & 0 & -\frac{c_{11}}{\rho} & -\frac{c_{12}}{\rho} &
    \frac{c_{13}}{\rho} &
    0 & 0 & 0\\
    0 & 0 & 0 & 0 & 0 & 0 & 0 & 0 & -\frac{c_{66}}{\rho} \\
    0 & 0 & 0 & 0 & 0 & 0 & 0 & -\frac{c_{55}}{\rho} & 0 \\
    -1 & 0 & 0 & 0 & 0 & 0 & 0 & 0 & 0 \\
    0 & 0 & 0 & 0 & 0 & 0 & 0 & 0 & 0 \\
    0 & 0 & 0 & 0 & 0 & 0 & 0 & 0 & 0 \\
    0 & 0 & 0 & 0 & 0 & 0 & 0 & 0 & 0 \\
    0 & 0 & -1 & 0 & 0 & 0 & 0 & 0 & 0 \\
    0 & -1 & 0 & 0 & 0 & 0 & 0 & 0 & 0 \\
  \end{pmatrix},\\
  \vec{A}_2
  =
  \begin{pmatrix}
    0 & 0 & 0 & 0 & 0 & 0 & 0 & 0 & -\frac{c_{66}}{\rho}\\
    0 & 0 & 0 & -\frac{c_{21}}{\rho} & -\frac{c_{22}}{\rho} &
    -\frac{c_{23}}{\rho}
    & 0 & 0 & 0 \\
    0 & 0 & 0 & 0 & 0 & 0 & -\frac{c_{44}}{\rho} & 0 & 0 \\
    0 & 0 & 0 & 0 & 0 & 0 & 0 & 0 & 0 \\
    0 & -1 & 0 & 0 & 0 & 0 & 0 & 0 & 0 \\
    0 & 0 & 0 & 0 & 0 & 0 & 0 & 0 & 0 \\
    0 & 0 & -1 & 0 & 0 & 0 & 0 & 0 & 0 \\
    0 & 0 & 0 & 0 & 0 & 0 & 0 & 0 & 0 \\
    -1 & 0 & 0 & 0 & 0 & 0 & 0 & 0 & 0 \\
  \end{pmatrix}, \\
  \vec{A}_3
  =
  \begin{pmatrix}
    0 & 0 & 0 & 0 & 0 & 0 & 0 & -\frac{c_{55}}{\rho} & 0\\
    0 & 0 & 0 & 0 & 0 & 0 & -\frac{c_{44}}{\rho} & 0 & 0 \\
    0 & 0 & 0 & -\frac{c_{31}}{\rho} & -\frac{c_{32}}{\rho} &
    -\frac{c_{33}}{\rho}
    & 0 & 0 & 0 \\
    0 & 0 & 0 & 0 & 0 & 0 & 0 & 0 & 0 \\
    0 & 0 & 0 & 0 & 0 & 0 & 0 & 0 & 0 \\
    0 & 0 & -1 & 0 & 0 & 0 & 0 & 0 & 0 \\
    0 & -1 & 0 & 0 & 0 & 0 & 0 & 0 & 0 \\
    -1 & 0 & 0 & 0 & 0 & 0 & 0 & 0 & 0 \\
    0 & 0 & 0 & 0 & 0 & 0 & 0 & 0 & 0 \\
  \end{pmatrix}.
  \label{eq:Aelasto}
\end{multline}
The energy functional for the elastodynamic wave problem reads:
\begin{equation}
  E_\text{elastic}(t)
  = \frac{1}{2}\int_{\Omega} \rho \vec{v} \cdot
  \vec{v} + \vec{\mathcal{C}^{-1}} \vec{T} \cdot \vec{T} \dif x.
\label{eq:elastodynamicenergy}
\end{equation}
The matrix $Q$ in~\eqref{eq:generalenergy} for this problem reads:
\begin{equation}
  Q = \text{diag}\del{\rho, \ \rho, \ \rho, \ \vec{C}^{-1}},
\label{eq:Qelasto}
\end{equation}
where $\vec{C}^{-1}$ is the stiffness tensor with in Voigt notation.
Note that the matrix $\vec{Q}$ is block diagonal for this case.

\section{Finite element formulations}
\label{sec:fem}

This appendix contains the finite element formulations for the
examples presented in Section~\ref{sec:results}. For the description
of the continuous problems we refer to
Section~\ref{sec:problem}. Unless mentioned otherwise in the text of
Section~\ref{sec:results}, linear elements are used in all cases.

\subsection{Acoustic wave propagation}
\label{sec:acdiscr}

We consider the acoustic wave problem described in
Section~\ref{sec:acoustic}. Applying the complex coordinate stretching
transformation~\eqref{eq:pml3d}, in the first dimension only,
to~\eqref{eq:asyst} we get the continuous system
\begin{equation}
  \begin{aligned}
    \frac{1}{K}\Dot{p} &=  -\nabla \cdot \vec{v} + \frac{\sigma_1}{K} p,
    \\
    \rho \Dot{\vec{v}} &=  -\nabla p + \vec{f} + \rho\sigma_1 \vec{v}.
  \end{aligned}
  \label{eq:amodel}
\end{equation}
The semi-discrete finite element problem of~\eqref{eq:amodel} reads:
find $\vec{v} \in \vec{U}$ and $p \in W$ such that
\begin{equation}
  \begin{aligned}
    \int_{\Omega} w \frac{1}{K} \Dot{p} \dif x
    &= -\int_{\Omega} w \nabla \cdot \vec{v} \dif x
    - \int_{\Omega} w \frac{\sigma_{1}}{K} p \dif x
    \quad \forall w \in W,
    \\
    \int_{\Omega} \vec{u} \cdot \rho \Dot{\vec{v}} \dif x
    & = \int_{\Omega} \nabla \cdot \vec{u} p \dif x
    - \int_{\Omega} \vec{u} \cdot \rho \sigma_{1}\vec{v} \dif x
    + \int_{\Omega} \vec{u} \cdot \vec{f} \dif x
    \quad \forall \vec{u} \in \vec{U},
  \end{aligned}
  \label{discretevaracosutic}
\end{equation}
where the function space $W \subset H^{1}_{0}(\Omega)$ is the usual
continuous Lagrange finite element space and $\vec{U} \subset
H_{0}({\rm div}, \Omega)$ is spanned by Raviart--Thomas
elements~\citep{raviart77thomas}. We use the same polynomial order for
both finite element spaces.  The classical leapfrog scheme is used to
advance in time.

\subsection{Elastic wave propagation}
\label{sec:eddiscr}

We will use the discontinuous Galerkin finite element method for the
elastodynamic example. Examples of a discontinuous Galerkin finite
element methods for elastic wave propagation can be found
in~\citep{boumatar12etall, etienne10etall}.

The semi-discrete discontinuous Galerkin finite element formulation~
for the two-dimensional generic PML description~\eqref{eq:2dfullpml}
on a triangulation $\mathcal{T} = \bigcup_{i=1}^{n_{k}} K_{i}$ of the
computational domain $\Omega$ into $n_{k}$ overlapping cells $K_{i}$,
reads: find $\vec{q} \in \vec{V}$ and $\vec{r} \in \vec{V}$ such that
for all triangles $K \in \mathcal{T}$
\begin{equation}
  \begin{aligned}
    \sum_{K} \int_{K_{i}} \del{\Dot{\vec{q}} + \sigma_{1+2}\vec{q} + \vec{r}
      - \vec{f}} \cdot \vec{l} \dif x
    - \sum_{K} \int_{K_{i}} \del{\vec{F}_{1}\cdot \vec{l}_{,1}
      + \vec{F}_{2} \cdot \vec{l}_{,2}} \dif x
\\ \quad \quad \quad \quad
    = - \sum_{K} \int_{\partial K_{i}} \del{n_1 \vec{F}_{1}^{*}+n_2
    \vec{F}_{2}^{*}}
    \cdot \vec{l} \dif s
    \quad \forall \vec{l} \in \vec{V},
    \\
    \sum_{K} \int_{K_{i}} \del{\Dot{\vec{r}}
      - \sigma_{12}\vec{q}} \cdot \vec{m} \dif x
    + \sum_{K} \int_{K_{i}} \del{\sigma_{2}\vec{F}_{1}\cdot \vec{m}_{,1}
      +  \sigma_{1}\vec{F}_{2} \cdot \vec{m}_{,2}} \dif x
\\ \quad \quad \quad \quad
    =
    \sum_{K} \int_{\partial K_{i}} \del{n_1
    \del{\sigma_{2}\vec{F}_{1}}^{*}+n_2
        \del{\sigma_{1}\vec{F}_{2}}^{*}} \cdot
    \vec{m} \dif s \quad \forall \vec{m} \in \vec{V},
  \end{aligned}
  \label{eq:globalweak}
\end{equation}
where the notation $\vec{l}_{,i} = \partial \vec{l}/ \partial x_{i}$
implies component-wise partial differentiation of $\vec{l}$ with
respect to~$x_{i}$, $\del{n_1,n_2}$ is the outward normal unit vector
to $\partial K$, $\vec{F}_{i}^{*}$ is the numerical flux (defined
below), $\vec{F}_{i} = \vec{A}_{i}\vec{q}$ and the used function space
is
\begin{equation}
  \vec{V}
  = \cbr{\vec{v} \in \sbr{L^{2}\del{\Omega}}^{5}
    : \vec{v}|_K \in \sbr{P_{k}\del{K}}^{5} \forall K \in \mathcal{T}},
\end{equation}
where $P_{k}\del{K}$ is the space of polynomial functions of degree $k
\geq 1$ on a cell~$K$. We choose Lax--Friedrichs numerical
fluxes~\cite[p.~34]{hesthaven08warburton}) to complete the
formulation:
\begin{equation}
    \vec{F}^{*}_i = \vec{A}_{i} \dfrac{\vec{q}^{+} + \vec{q}^{-}}{2} +
    \frac{C}{2}\del{n_i^{+} \vec{q}^{+} + n_{i}^{-} \vec{q}^{-}},
    \qquad
    \del{\sigma_j\vec{F}_i}^{*} = \vec{A}_{i} \dfrac{\sigma_j^{+}\vec{q}^{+}
    + \sigma_j^{-}\vec{q}^{-}}{2} +
    \frac{C}{2}\del{n_i^{+} \vec{q}^{+} + n_{i}^{-}
    \vec{q}^{-}}.
    \label{eq:numflux}
\end{equation}
where the `$+$' and `$-$' superscripts indicate the interior and
exterior side of an interface and $C$ is the greatest wave speed
occurring in the problem. The trapezoidal rule is used to advance in
time. All boundary conditions are enforced weakly.

\subsection{Electromagnetic wave propagation}
\label{sec:em_discr}

The semi-discrete finite element formulation for the electromagnetic
wave propagation problem reads: find $\vec{H} \in \vec{U} \subset
H({\rm div}, \Omega)$ and $\vec{E} \in \vec{V} \subset H({\rm
  curl}, \Omega)$ such that
\begin{equation}
  \begin{aligned}
    \int_{\Omega} \vec{u} \cdot \mu \Dot{\vec{H}}
    &= -\int_{\Omega}\vec{u} \cdot \nabla \times \vec{E} \dif x
    - \int_{\Omega}\vec{u} \cdot \mu\sigma \vec{H} \dif x
    \quad \forall \vec{u} \in \vec{U},
    \\
    \int_{\Omega} \vec{v} \cdot \varepsilon \Dot{\vec{E}}
    &= \int_{\Omega} \nabla \times \vec{v} \cdot \vec{H} \dif x
    - \int_{\Omega} \vec{v} \cdot \varepsilon \sigma \vec{E} \dif x
    \quad \forall \vec{v} \in \vec{V},
  \end{aligned}
  \label{eq:emdiscr}
\end{equation}
where the function spaces $\vec{U}$ and $\vec{V}$ are spanned by
Raviart--Thomas elements~\citep{raviart77thomas} and N\'ed\'elec
elements of the first kind~\citep{nedelec80}, respectively. We use the
same polynomial order for both kinds of the elements.  The Yee
scheme~\citep{yee66} is used to advance in time.

\section{Computing the gradient of the objective functional for the
  time discretised problem}
\label{sec:gradient}

Consider the generic wave equation with a PML in one spatial dimension
as given in~\eqref{eq:1dfullpml} to be discretised in time with the
implicit trapezoidal rule, with one constant matched layer added:
\begin{equation}
  \frac{\vec{q}_{n+1} - \vec{q}_{n}}{\Delta t}
  + \vec{A}_{1}\del{\frac{\vec{q}_{n + 1}^\prime + \vec{q}_{n}^\prime}{2}}
  + \sigma \frac{\vec{q}_{n + 1}+\vec{q}_{n}}{2} = \vec{f},
\end{equation}
where $\vec{q}_{n}$ is the computed approximation for
$\vec{q}\del{n\Delta t}$ and the accent indicates a spatial
derivative, i.e., $\vec{q}_{n}^\prime = \partial\vec{q}_n/\partial
x_1$.  We introduce the vector $\Bar{\vec{q}} = \del{\vec{q}_{0}, \,
  \vec{q}_{1}, \, \ldots, \, \vec{q}_{n-1}, \, \vec{q}_{n}}$
containing the solution at each time step.  To be able to study the
procedure in detail, we will restrict the time integration to two
steps.  In that case the objective function is
\begin{equation}
  J(\Bar{\vec{q}}, \vec{u})
  = \frac{1}{2} \left<\vec{Q} \vec{q}(t), \vec{q}(t) \right>_\Omega,
\end{equation}
and its derivative with respect to the state vector $\Bar{\vec{q}}$ is
\begin{equation}
\pd{J}{\Bar{\vec{q}}}
=
\begin{pmatrix}
0 \\ 0 \\ \vec{Q}\vec{q}_{2}
\end{pmatrix}.
\end{equation}
The system of constraints in this case consists of three equations:
\begin{equation}
\begin{aligned}
\vec{c}_{0} &= \vec{q}_{0} - \vec{f}_{0} = \vec{0},
\\
\vec{c}_{1} &= \frac{\vec{q}_{1} - \vec{q}_{0}}{\Delta t}
+ \vec{A}_{1}\del{\frac{\vec{q}_{1}^\prime + \vec{q}_{0}^\prime}{2}}
+ \sigma \frac{\vec{q}_{1}+\vec{q}_{0}}{2} = 0,
\\
\vec{c}_{2} &= \frac{\vec{q}_{2} - \vec{q}_{1}}{\Delta t}
+ \vec{A}_1 \del{\frac{\vec{q}_{2}^\prime + \vec{q}_{1}^\prime}{2}}
+ \sigma \frac{\vec{q}_{2} + \vec{q}_{1}}{2} = 0,
\end{aligned}
\label{eq:forwardsolve}
\end{equation}
for which we can compute the Jacobian matrix
\begin{equation}
  \dpd{\vec{c}}{\Bar{\vec{q}}}
  =
  \begin{pmatrix}
    \pd{\vec{c}_{0}}{\vec{q}_{0}}
    & \pd{\vec{c}_{1}}{\vec{q}_{0}}
    & \pd{\vec{c}_{2}}{\vec{q}_{0}}
    \\
    \pd{\vec{c}_{0}}{\vec{q}_{1}}
    & \pd{\vec{c}_{1}}{\vec{q}_{1}}
    & \pd{\vec{c}_{2}}{\vec{q}_{1}}
    \\
    \pd{\vec{c}_{0}}{\vec{q}_{2}}
    & \pd{\vec{c}_{1}}{\vec{q}_{2}}
    & \pd{\vec{c}_{2}}{\vec{q}_{2}}
  \end{pmatrix}
  =
  \begin{pmatrix}
    \vec{I} & \pd{\vec{c}_{1}}{\vec{q}_{0}} & 0
    \\
    0 & \pd{\vec{c}_{1}}{\vec{q}_{1}} & \pd{\vec{c}_{2}}{\vec{q}_{1}}
    \\
    0 & 0 & \pd{\vec{c}_{2}}{\vec{q}_{2}}
  \end{pmatrix},
\end{equation}
where $\vec{I}$ is the identity matrix of size $n\times n$ where
$n$ is the length of $\vec{q}_i$. To determine this matrix, four
non-trivial partial derivatives have to be computed. In general, all
we need is
\begin{equation}
  \begin{aligned}
    \dpd{\vec{c}_{i}}{\vec{q}_{i}}
    &= \frac{\vec{I}}{\Delta t}
    + \vec{A}_{1} \del{\frac{\vec{D}}{2}}
    + \sigma \frac{\vec{I}}{2},
    \\
    \dpd{\vec{c_i}}{\vec{q}_{i-1}}
    &= -\frac{\vec{I}}{\Delta t}
    + \vec{A}_{1} \del{\frac{\vec{D}}{2}}
    + \sigma \frac{\vec{I}}{2},
  \end{aligned}
\end{equation}
where $\vec{D}$ is a diagonal matrix with $\vec{D}_{ii} = \partial_{x_{1}}$
This information allows us to compute the adjoint states
$\vec{\lambda}$ by solving
\begin{equation}
  \pd{\vec{c}}{\Bar{\vec{q}}} \vec{\lambda} = -\pd{J}{\Bar{\vec{q}}},
\end{equation}
which in this case looks like
\begin{equation}
  \begin{pmatrix}
    \vec{I} & \pd{\vec{c}_{1}}{\vec{q}_{0}} & 0
    \\
    0 & \pd{\vec{c}_{1}}{\vec{q}_{1}} & \pd{\vec{c}_{2}}{\vec{q}_{1}}
    \\
    0 & 0 & \pd{\vec{c}_{2}}{\vec{q}_{2}}
  \end{pmatrix}
  \begin{pmatrix}
    \vec{\lambda}_{0} \\ \vec{\lambda}_{1} \\ \vec{\lambda}_{2}
  \end{pmatrix}
  =-
  \begin{pmatrix}
    0 \\ 0 \\ \vec{q}_{2}
  \end{pmatrix}.
\end{equation}
Solving this system by back substitution leads to
\begin{equation}
  \begin{aligned}
    0 &= \vec{\lambda}_{2} + \vec{q}_{2},
    \\
    0 &= \frac{\vec{\lambda}_{1} - \vec{\lambda}_{2}}{\Delta t}
    + \vec{A}_1\del{\frac{\vec{\lambda}_{1}^\prime
        + \vec{\lambda}_{2}^\prime}{2}}
    + \sigma \frac{\vec{\lambda}_{1}+\vec{\lambda}_{2}}{2},
    \\
    0 &= \frac{\vec{\lambda}_{0} - \vec{\lambda}_{1}}{\Delta t}
    + \vec{A}_1\del{\frac{\vec{\lambda}_{0}^\prime +
    \vec{\lambda}_{1}^\prime}{2}}
    + \sigma \frac{\vec{\lambda}_{0}+\vec{\lambda}_{1}}{2},
  \end{aligned}
  \label{eq:adjointsolve}
\end{equation}
which is identical to solving the forward
problem~\eqref{eq:forwardsolve} up to the variable names. In fact, by
feeding $-\vec{q}_{2}$ to the forward solver as a source, the adjoint
states will be computed in reversed order.

To obtain the gradient, all that remains is to multiply the forward
and adjoint states:
\begin{equation}
  \od{J}{\vec{u}} = \sum_{i = 0}^{n} \left<\vec{\lambda}_{i}^T,
  \vec{q}_{i}\right>.
\end{equation}

\end{document}